\documentclass[12pt]{article}

\usepackage{amsmath}
\usepackage{amssymb}
\usepackage{indentfirst}
\usepackage{amssymb}
\usepackage{amsfonts}
\usepackage{amscd}
\usepackage{amsbsy}
\usepackage{amsthm}
\usepackage{latexsym}
\usepackage{graphicx,color} 
\usepackage[dvipsnames]{xcolor}
\usepackage[colorlinks]{hyperref}
\hypersetup{linkcolor=blue,citecolor=blue,urlcolor=blue}

\def\nn{\nonumber}       
\def\beq{\begin{eqnarray}}
\def\eeq{\end{eqnarray}}

\def\ln{\,\mbox{ln}\,}

\def\al{\alpha}
\def\be{\beta}

\def\ga{\gamma}
\def\de{\delta}
\def\vp{\varepsilon}
\def\ep{\epsilon}

\def\la{\lambda}
\def\na{\nabla}
\def\pa{\partial}

\def\si{\sigma}
\def\om{\omega}
\def\ph{\varphi}

\def\th{\theta}

\def\Ga{\Gamma}
\def\De{\Delta}
\def\La{\Lambda}

\DeclareMathOperator{\cx}{\square}
\def\BV{Batalin-Vilkovisky }

\begin{document}

\begin{center}

{\Large
Gauge invariant renormalizability of quantum gravity}\footnote{
Invited chapter for the Section "Effective Quantum Gravity"
of the "Handbook of Quantum Gravity" (Eds. C. Bambi, L. Modesto
and I.L. Shapiro, Springer Singapore, expected in 2023).}
\vskip 3mm

\textbf{P.M. Lavrov}$^{(a)}$\footnote{E-mail address: lavrov@tspu.edu.ru}
\ \
and
\ \ \textbf{I.L. Shapiro}$^{(b)}$\footnote{E-mail address:
ilyashapiro2003@ufjf.br
, \ \
On leave from Tomsk State Pedagogical University, Russia}

\vskip 4mm

(a) \ \ Center of Theoretical Physics, Tomsk State
Pedagogical University
\\
Kievskaya St. 60, 634061, Tomsk, Russia
\vskip 2mm

(b) \ \ Departamento de F\'{\i}sica,
Universidade Federal de Juiz de Fora,
\\
Juiz de Fora, MG, 36036-900, Brazil

\end{center}

%


\begin{quotation}

\noindent
\textbf{Abstract.}
The current understanding of renormalization in quantum gravity (QG)
is based on the fact that UV divergences of effective actions in the
covariant QG models are covariant local expressions. This fundamental
statement plays a central role in QG and, therefore, it is important to
prove it for the widest possible range of the QG theories. Using the
Batalin-Vilkovisky technique and the background field method, we
elaborate the proof of gauge invariant renormalizability for a generic
model of quantum gravity that is diffeomorphism invariant and does
not have additional, potentially anomalous, symmetries.
\vskip 2mm

\vskip 2mm


\end{quotation}

\section{Introduction}
\label{secI}

All the fundamental interactions that exist in Nature (electroweak,
strong and gravitational) are described in terms of gauge theories
\cite{Weinberg}. Thus, the quantization of gauge theories that
provides important insights into the quantum properties of the
fundamental forces plays an important role in this process.
General relativity (GR) and most of its extensions are based
on the general covariance principle. In the models where this
symmetry does not hold at the classical level (let us note that there
is a special Chapter in this Section, devoted to the Ho\v{r}ava, or
Ho\v{r}ava - Lifshitz, gravity models, in which the covariance is
violated), this symmetry is supposed to restore in the limit
corresponding to observations. Thus, the basic models of QG have
to be based on the covariant classical theory.

At the classical level, the coordinate transformations change the
metric (or other fields, describing gravity, e.g., independent
connection, or tetrad, etc) to the equivalent one, from a physical
viewpoint. The situation is the same as in other gauge theories, with
the coordinate invariance playing the role of gauge symmetry. In this
review, we discuss only the QG models in which gravity is described
by the metric. Thus, we are leaving aside such interesting issues as
first-order (Palatini) formalism in QG \cite{Popov-FI}, teleparallel
gravity \cite{AldroPer}, etc. Let us note, by passing, that the
quantum aspects of the teleparallel gravity are not well elaborated.

In quantum theory, the gauge symmetry results in degeneracy and
the difficulties in the immediate application of quantum mechanics,
in both Hamiltonian (canonical) and the Lagrangian (covariant)
formalisms. In this Section, we concentrate our attention on the
Lagrangian approach and the path integral technique, which is an
essential ingredient of covariant quantization and provides economic
way of getting the Feynman rules directly from the classical
Lagrangian.

The relevant question is what happens with the diffeomorphism
invariance at the quantum level. In particular, is it true that this
symmetry holds after the loop corrections are taken into account?
Or, in a more particular way, does the general covariance hold in
the counterterms that cancel the divergences of the effective action
in the model of QG? In what follows, we shall see that the answer
to both questions may be positive for a variety of the QG models.

Since QG represents a particular type of gauge theories, the
quantization of gravity should follow the same rules as in other
gauge theories. The covariant quantization of gauge theories has
made a long way starting from the famous work of Feynman \cite{F}
where the non-unitarity of $S$-matrix in Yang-Mills \cite{YM} and
gravity theories have been found within the approach based on a naive
quantization procedure. This procedure was developed in quantum
electrodynamics (QED), which is an Abelian
gauge theory. Later on, consistent covariant quantization of the
Yang-Mills theories has been found in the papers by Faddeev and
Popov \cite{FP}, and DeWitt \cite{DeW}. The functional integration
measure that appears in the course of quantization \cite{DeW}, can be
encoded into an additional functional integration over the so-called
ghost fields or Faddeev-Popov ghosts \cite{FP}. The Faddeev-Popov
method operates with the complete (also called total) quantum action
which is a sum of the initial classical action, the ghost action, and the
gauge fixing action. The total action should be used for the
construction of generating functional of Green functions.

The first proof of the gauge invariant renormalizability in the
Yang-Mills theories has been given by 't Hooft and Veltman in
\cite{tHooft-71,Hooft_Veltman-NP}, that was an important
contribution to creating the quantum field theory (QFT)
foundations of the Standard Model of particle physics and its
extensions. These works have been awarded the 1999 Nobel Prize
in Physics ``for elucidating the quantum
structure of electroweak interactions in physics''. The proof of
renormalizability was based on the basic Ward identities, which was
a very complicated approach. Soon after that, there was developed
new powerful techniques of exploring non-Abelian gauge symmetries
at the quantum level. One of the most remarkable discoveries in this
respect was that the total Faddeev-Popov-DeWitt action is invariant
under global supersymmetry called the BRST symmetry (named by
Becchi, Rouet, Stora, and Tyutin) \cite{BRS1,BRS,T}).
The nilpotent BRST charge plays a crucial role in the construction
of the physical state space of any Yang-Mills - type theory allowing
to define the gauge-independent and unitary $S$-matrix \cite{KO}.

The most efficient way to explore the gauge symmetry at the quantum
level is the Batalin-Vilkovisky technique, that is a natural
extension of the DeWitt -- Faddeev-Popov \cite{DeW,FP} procedure
and the Becchi-Rouet-Stora and Tyutin (BRST) symmetry \cite{BRS,T}.
The Batalin-Vilkovisky approach can be most successfully applied in
the theory of the Yang-Mills type when the generators of gauge
transformations form a closed algebra. We shall see, in the next
sections, that QG, with the diffeomorphism invariance as a gauge
symmetry, belongs to this class of theories.  Another  relevant
restriction is that there should not be other symmetries that can
``compete'' with the diffeomorphism invariance at the quantum
level. A simple criterium to see whether an additional symmetry is
really spoiling the renormalizability is to check whether there is a
regularization preserving both symmetries at the same time.
In all known cases, the lack of such symmetry means there are
quantum anomalies. A typical example is the local conformal
symmetry or the Weyl symmetry. It is expected that the anomaly
violates the renormalizability of the theories of QG (contrary to
the QFT in curved spacetime) with this symmetry at higher loop
orders. However, since this is an advanced (and not very
well-elaborated subject, we shall not be concerned about it in
the present review.
Thus, in what follows we assume the use of dimensional
regularization and the absence of conformal symmetry.

The  Batalin-Vilkovisky technique implies introducing
additional objects called antifields. With antifields, the BRST
symmetry and the corresponding equations for the effective action
(Zinn-Justin equation \cite{Z-J} and Slavnov-Taylor identities in
Yang-Mills theory \cite{tHooft-71,Slavnov,Taylor}) become more
straightforward to analyze. The gauge invariant renormalizability
guarantees that in all orders of loop expansion for the quantum
effective action, one can control deformations of the generators
of gauge transformations, which leave such an action invariant.
In the background field method, one can maintain general covariance
of the divergent part of effective action when the mean quantum
fields, ghosts, and antifields are switched off.

The traditional (and correct) view of the difficulty in quantizing
the gravitational field is that the quantum GR is not a renormalizable
theory, while the renormalizable version of the theory includes
fourth derivatives in the action \cite{Stelle77} and therefore it is
not unitary. In the last decades, this simple two-side story was
getting more complicated, with the new models of superrenormalizable
gravity, both polynomial \cite{ highderi} and non-polynomial
\cite{krasnikov,kuzmin,Tomboulis-97}. Typically, these models
intend to resolve the conflict between non-renormalizability and
non-unitarity by introducing the finite or infinite amounts of
derivatives compared to the fourth-derivative model \cite{Stelle77}.

The main advantage of the non-polynomial models is that the tree
level propagator  may have the unique physical pole corresponding
to the massless graviton. At the same time, the dressed propagator has,
typically, an infinite (countable) amount of the ghost-like states
with complex poles \cite{CountGhost} and hence the questions about
physical contents and quantum consistency of such a theory remains
open, especially taking into account the problems with reflection
positivity \cite{ARS,ChrMod}.

On the other hand, within the polynomial model, one can prove the
unitarity of the $S$-matrix within the Lee-Wick approach \cite{LW}
to quantum gravity
in four  \cite{LM-Sh} and even higher dimensional space-times
\cite{LWQG}. Furthermore, it is possible to make explicit one-loop
calculations \cite{MRS} which provide exact beta-functions
in these theories due to the superrenormalizability of the theory.
In the part of stability, the existing investigations concerned
special backgrounds, namely cosmological \cite{HD-Stab,PeSaSh}
and black hole cases \cite{Whitt,Myung,Aux}. While the black
hole results do not look conclusive, the results for the cosmological
backgrounds provide a good intuitive understanding of  the problem
of stability in the gravity models with higher derivative ghosts.

Independent on the efforts in better understanding the role of ghosts
and instabilities in both polynomial and non-polynomial models, it
would be useful to have formal proof that, at the quantum level,
these theories are renormalizable or superrenormalizable. The first
proof was given for the  fourth derivative quantum gravity
\cite{Stelle77},
with some simplifications and generalizations  achieved in
\cite{VorTyu82,VorTyu84} as an application of a general approach
\cite{VorTyu-gen-1982}. Recently, there was a renewed interest
in this subject, in particular, the proofs of the general-covariant
renormalizability in the general types of local \cite{Barv} models
of QG and even a more general Batalin-Vilkovisky-based proof
of a gauge invariant
renormalizability in the general models of quantum gravity, which
may include higher derivatives, including polynomial or
non-polynomial models \cite{Lavrov-renQG}. In what follows,
we shall closely follow the last reference, just adding a few more
details.

The review is organized as follows. In Sec.~\ref{sec2}, we formulate
the Batalin-Vilkovisky formalism combined with the background field
method in QG. In Sec.~\ref{sec3}, this formalism is applied to the
formal proof of renormalizability in the model of QG of the general
form. On top of that, we discuss the gauge fixing independence of
the vacuum functional and of the gravitational $S$-matrix.
Sec.~\ref{sec4} contains the short historic note, the extended list
of the main references (which certainly remains incomplete owing
to the limited size of the review) and a brief discussion of the
requirements on the gauge theory which enable one to use the
Batalin-Vilkovisky formalism. In Sec.~\ref{sec5}, the gauge fixing
in the higher derivative models is discussed.
Finally, in Sec.~\ref{secC}, we draw our conclusions.

Condensed DeWitt's notations \cite{DeWitt} are used in the review.
Right and left derivatives of a quantity $f$ with respect to the
variable $\varphi$ are denoted as $\frac{\de_r f}{\de \ph}$ and
$\frac{\de_l f}{\de \ph}$, correspondingly. The Grassmann parity
and the ghost number of a quantity $A$ are denoted by $\vp(A)$ and
${\rm gh}(A)$, respectively. The reader can see Eq.~(\ref{GN}) for
the definition, in the last case. For the sake of generality, we
perform all general considerations for an arbitrary spacetime
dimension $D$.
Let us note that this is different from the previous Chapter of this
Section of the Handbook, where it was assumed $D=4$. The
condensed notation for the space-time integral in $D$ dimensions,
$\int dx =\int d^Dx$ is used throughout the text of the present
Chapter.

\section{Quantum gravity in the background field formalism}
\label{sec2}

Our purpose is to explore the gravitational theory based on an
arbitrary action of a Riemann's metric $\,S_0(g)$, where
$\,g = g_{\mu\nu}(x)$. In what follows we usually omit the
indices in the arguments of functions or functionals. The action
is assumed invariant under the general coordinate transformations,
\beq
\label{A1}
x^{\prime\,\mu} = x^{\prime\,\mu}(x)
,\qquad
g'_{\mu\nu}(x') \,=\,
\frac{\pa x^{\alpha}}{\pa {x'}^{\mu}}
\frac{\pa x^{\beta}}{\pa {x'}^{\nu}}
\,g_{\alpha\beta}(x).
\eeq

The standard examples of the theories of our interest are
Einstein's gravity (GR) with a cosmological constant term,
\beq
S_{EH}(g)
&=& - \, \frac{1}{\kappa^2}
\int dx \sqrt{-g}\, \big(R + 2\La\big)
\label{EH}
\eeq
and a general version of higher derivative gravity. The corresponding
action is described in the previous Chapter \cite{BackQG},
\beq
S(g)  &\,=\,&
S_{EH}(g)
\,+\,
\int dx \sqrt{-g}\,\bigg\{
R^{\mu\nu\al\be} \Pi_1\Big( \frac{\square}{M^2} \Big) R_{\mu\nu\al\be}
\nn
\\
&&
+\,\, R^{\mu\nu} \Pi_2 \Big( \frac{\square}{M^2} \Big) R_{\mu\nu}
\,+ \, R\Pi_3 \Big( \frac{\square}{M^2} \Big) R
\,\,+ \,\, {\cal O} \big( R_{...}^3\big)\bigg\}.
\label{act1}
\eeq
Here $\Pi_{1,2,3}$ are polynomial or non-polynomial form factors
in the quadratic in curvatures part of the Lagrangian and the last
term represents non-quadratic in curvature terms. In quantum theory,
action
(\ref{act1}) may lead to the theory which is non-renormalizable,
renormalizable or superrenormalizable, depending on the choice
of the functions $\Pi_{1,2,3}(x)$ and the non-quadratic terms.

The dimensional parameter $M^2$ in the form factors
$\,\Pi_{1,2,3}\,$ is a universal mass scale at
which the QG effect becomes relevant. For instance, $M^2$  can
be the square of the Planck mass, but there may be other options,
including multiple scale models, as analyzed in \cite{ABS-SeeSaw}.
For the analysis presented below, these (otherwise important)
details are irrelevant since the unique necessary feature is that
the action $S(g)$ is diffeomorphism invariant.

In the infinitesimal form, the transformations (\ref{A1}) read
\beq
\label{A2}
{x'}^{\mu}=x^{\mu}+\xi^{\mu}(x)
\quad
\Longrightarrow
\quad
g'_{\mu\nu}(x)=g_{\mu\nu}(x)+\delta g_{\mu\nu}(x),
\eeq
where
\beq
\label{A3}
\delta g_{\mu\nu}(x)=-\xi^{\sigma}(x)\pa_{\sigma}g_{\mu\nu}(x)
- g_{\mu\si}(x)\pa_{\nu}\xi^{\si}(x)
- g_{\si\nu}(x)\pa_{\mu}\xi^{\si}(x).
\eeq
On top of (\ref{A3}), we also need the transformation rule for
vector fields $A_{\mu}(x)$ and $A^{\mu}(x)$,
\beq
\label{A5}
&&
\delta A_{\mu}(x)
\,=\,-\,\xi^{\sigma}(x)\pa_{\sigma}A_{\mu}(x)-
A_{\sigma}(x)\pa_{\mu}\xi^{\sigma}(x),
\\
\label{A6}
&&
\delta A^{\mu}(x)
\,=\,-\,\xi^{\sigma}(x)\pa_{\sigma}A^{\mu}(x)+
A^{\sigma}(x)\pa_{\sigma}\xi^{\mu}(x).
\eeq

The invariance of the action $\,S_0(g)\,$ under the transformations
\ (\ref{A3}) \ can be expressed in the form of Noether identity
\beq
\label{A4}
\int dx \,\,\frac{\delta S_0(g)}{\delta g_{\mu\nu}(x)}
\,\,\delta g_{\mu\nu}(x)\,=\,0.
\eeq

Let us present the transformations (\ref{A3}) in the form
\beq
\label{A5a}
\delta g_{\mu\nu}(x)
\,=\,\int dy\, \,R_{\mu\nu\sigma}(x,y;g)\,\xi^{\sigma}(y),
\eeq
where the generators of gauge transformations of the metric tensor
$g_{\mu\nu}$ with gauge parameters $\xi^{\sigma}(x)$ are
\beq
\label{A6b}
R_{\mu\nu\sigma}(x,y;g)
&=&
- \delta(x-y)\pa_{\sigma}g_{\mu\nu}(x)
- g_{\mu\sigma}(x)\pa_{\nu}\delta(x-y)
\nn
\\
&&
-\,\, g_{\sigma\nu}(x)\pa_{\mu}\delta(x-y).
\eeq
The algebra of generators (also called  algebra of gauge
transformations), can be easily verified to be
\beq
&&
\int du \,\bigg[
\frac{\de R_{\mu\nu\si}(x,y;g)}{\de g_{\al\be}(u)}
\,R_{\alpha\beta\ga} (u,z;g)
\,\,-\,\,
\frac{\de R_{\mu\nu\ga}(x,z;g)}{\de g_{\al\be}(u)}
\,R_{\alpha\beta\sigma} (u,y;g)\bigg]
\nn
\\
\label{A7}
&&
\qquad\qquad\qquad
= \,\, -\,
\int du \,\,
R_{\mu\nu\la}(x,u;g)\,F^{\la}_{\si \ga}(u,y,z),
\eeq
where
\beq
F^{\lambda}_{\alpha\beta}(x,y,z)
&=&
\delta (x-y)\;\delta^{\lambda}_{\beta}\;\frac{\pa }{\pa x^{\alpha}}
\delta(x-z)
\,-\, \delta(x-z)\;\delta^{\lambda}_{\alpha}\;\frac{\pa }{\pa x^{\beta}}
\delta (x-y),
\quad
\nn
\\
F^{\lambda}_{\alpha\beta}(x,y,z)
&=&-\,F^{\lambda}_{\beta\alpha}(x,z,y)
\label{A8a}
\eeq
are structure functions of the gauge algebra which do not depend
on the metric $g_{\mu\nu}$.

The algebra (\ref{A7}) holds independently of the form of the
covariant action. Thus, any covariant theory of gravity is a
gauge theory with closed gauge algebra and with the structure
functions independent of the fields (metric tensor, in this case).
Thus, QG is a theory of the Yang-Mills type.

Starting from this point, the analysis of renormalization is pretty
much standard, but it is important to introduce one more important
element in our considerations. It proves useful to perform
quantization of gravity not on the Minkowski space-time with
the constant metric tensor $\eta_{\mu\nu}$, but on undefined
external background, represented by the metric tensor
${\bar g}_{\mu\nu}(x)$.  Introducing an arbitrary background
metric provides serious advantages by using the \textit{background
field method}, as we shall see in what follows.
The standard references on the background field method in
QFT are \cite{DeW,AFS,Abbott}. One can see also recent
discussions of the method for the gauge theories in
\cite{BLT-YM,FT,Lav,BLT-YM2}) and specifically for QG
in  \cite{Barv}.

In the background field method, the metric $g_{\mu\nu}(x)$
is replaced by the sum
\beq
g_{\mu\nu}(x)
\,\,\, \longrightarrow \,\,\,
{\bar g}_{\mu\nu}(x)+h_{\mu\nu}(x),
\quad\,\,
\mbox{such that}
\qquad
S_0(g)
\,\,\, \longrightarrow \,\,\,
S_0({\bar g}+h).
\quad
\label{A9}
\eeq
Here $h_{\mu\nu}(x)$ is called quantum metric and is regarded
as an integration variable in the functional integrals defining the
generating functionals of Green functions.

The action $S_0({\bar g}+h)$ is a functional of two variables
${\bar g}$ and $h$ and has an additional symmetry corresponding
to the transformations
\beq
\delta{\bar g}_{\mu\nu}=\epsilon_{\mu\nu}
\qquad
\mbox{and}
\qquad
\delta h_{\mu\nu}=-\epsilon_{\mu\nu},
\label{eps}
\eeq
with arbitrary symmetric tensor functions
$\ep_{\nu\mu}=\ep_{\mu\nu}=\epsilon_{\mu\nu}(x)$. In particular,
this implies an ambiguity in defining the gauge transformations for
${\bar g}$ and $h$.
To fix this arbitrariness, we require that the transformation of
our interest has the right flat limit when ${\bar g}_{\mu\nu}(x)$ is
replaced by $\eta_{\mu\nu}$. In this way, one can easily show that
the gauge transformation of the quantum metric $h_{\mu\nu}$, in
the presence of external (fixed) background ${\bar g}_{\mu\nu}$,
has the form
\beq
\delta h_{\mu\nu }(x)\,=\,
\int dy \,R_{\mu\nu\sigma}(x,y;{\bar g}+h)\xi^{\sigma},
\label{A10}
\eeq
while $\de {\bar g}_{\mu\nu}(x) = 0$ and the action remains
invariant, $\delta {S}_{0}({\bar g}+h)=0$.

Because of the similarity with the Yang-Mills field, the
Faddeev-Popov quantization procedure is quite standard (see, e.g.,
the previous Chapter of this Section for the specific QG details)
and the resulting  action $S_{FP}=S_{FP}(\phi,{\bar g})$ has the
form \cite{FP}
\beq
\label{A11}
S_{FP}\,=\,
S_{0}({\bar g}+h)+S_{gh}(\phi, {\bar g})+S_{gf}(\phi,{\bar g}),
\eeq
where the contents of $\phi$ depends on the choice of the QG model
(\ref{act1}), as explained below. In the presence of a background
metric, the ghost action has the form
\beq
\label{A11a}
S_{gh}(\phi,{\bar g})\,=\,
\int dx dydz \sqrt{-{\bar g}(x)}\;
{\bar C}^{\alpha}(x) \,
H_{\alpha}^{\beta\gamma}(x,y;{\bar g},h)\,
R_{\beta\gamma\sigma}(y,z;{\bar g}+h)\,C^{\sigma}(z),
\quad
\eeq
with  the notation
\beq
\label{A11b}
H_{\alpha}^{\beta\ga}(x,y;{\bar g},h)=
\frac{\delta\chi_{\al}(x;{\bar g},h)}{\delta h_{\be\ga}(y)}.
\eeq
Furthermore, $S_{gf}({\bar g},h)$ is the gauge fixing action
\beq
\label{A11c}
S_{gf}(\phi,{\bar g})=\int dx \sqrt{-{\bar g}(x)}\;
B^{\alpha}(x)\chi_{\alpha}(x;{\bar g},h).
\eeq
which corresponds to the singular gauge condition. For the
non-singular gauge condition the action has the form
\beq
S_{gf}(\phi,{\bar g})=\int dx \sqrt{-{\bar g}(x)}\;
\Big[B^{\alpha}(x)\chi_{\alpha}(x;{\bar g},h)+\frac{1}{2}B^\al(x)
{\bar g}_{\alpha\beta}(x)B^{\beta}(x)\Big].
\label{A11d}
\eeq
The reader can easily identify (\ref{A11d}) as the particular
version of the general gauge fixing choice discussed in the
previous Chapter \cite{BackQG} and also note that (\ref{A11c})
is a degenerate version of
(\ref{A11d}). In what follows, we shall use the form (\ref{A11c}),
which is most useful for exploring the general features of
renormalization. Finally, $\chi_{\alpha}(x;{\bar g},h)$ are the gauge
fixing functions. It is worth mentioning that the generalization to a
more complicated non-singular gauge fixing, required in the higher
derivative theories, is considered in Sec.~\ref{sec5}.

Now we are in a position to introduce an important notation used
in (\ref{A11}),
\beq
\phi\, =\,\{\phi^i\}
\,\,=\,\, \big\{h_{\mu\nu},\,B^\al,\,C^\al,\,{\bar C}^\al\big\},
\label{GP}
\eeq
which is the full set of quantum fields including quantum metric,
Faddeev-Popov ghost, anti-ghost and the Nakanishi-Lautrup
auxiliary fields $B^{\alpha}$.

We will need two sets of numbers characterizing the quantum fields.
The Grassmann parity of the fields $\phi^i$ will be denoted as
\ $\vp(\phi^i)=\vp_i$. By definition, for ghost and anti-ghost we
have \ $\vp(C^{\alpha})=\vp( {\bar C}^{\alpha})=1$, while
for the 
auxiliary fields $B^{\alpha}$ and the quantum metric we have
\ $\vp(B^{\alpha})=\vp(h_{\mu\nu})=0$.

On top of this, another conserved quantity called ghost number,
can be defined for the same fields as
\beq
{\rm gh}(C^{\alpha}) = 1,
\qquad
{\rm gh}({\bar C}^{\alpha})=-1
\qquad
\mbox{and}
\qquad
{\rm gh}(B^{\alpha})={\rm gh}(h_{\mu\nu})=0.
\label{GN}
\eeq

Consider the gravitational BRST transformations, which were
introduced in \cite{DR-M,Stelle77,TvN}. For any admissible choice
of gauge fixing functions $\chi_{\alpha}(x;{\bar g},h)$ action
(\ref{A11}) is invariant under global supersymmetry (BRST
symmetry) \cite{BRS1,T},
\beq
\nonumber
&&
\delta_B h_{\mu\nu}(x)
 = \int dy R_{\mu\nu\alpha}(x,y;{\bar g}+h)C^\al(y)\mu,
\nn
\\
&&
\delta_B B^{\alpha}(x) = 0,
\nn
\\
\label{A15}
&&
\delta_B C^{\alpha}(x) = -\,C^{\sigma}(x)\pa_{\sigma}
C^{\alpha}(x)\mu,
\\
&&
\delta_B {\bar C}^{\alpha}(x)=B^{\alpha}(x)\mu,
\nn
\eeq
where $\mu$ is a constant Grassmann parameter, such that $\mu^2=0$.
It is a remarkable fact that the BRST transformations (\ref{A15}) do
not depend on gauge fixing condition. One can present the
BRST transformations (\ref{A15}) in the form
 \beq
\label{A16}
\delta_{B}\phi^i(x) \,=\, R^i(x;\phi, {\bar g})\,\mu,
\eeq
where
$R^i = \big\{ R^{(h)}_{\mu\nu},\, R_{(B)}^{\,\al},\,R_{(C)}^{\,\al},\,
R_{({\bar C})}^{\,\al}\big\}$ and
\beq
\label{A17}
&&
R^{(h)}_{\,\mu\nu}(x; \phi,{\bar g})
\,=\,
\int dy\, R_{\mu\nu\sigma}(x,y;{\bar g}+h)\,C^{\sigma}(y),
\nn
\\
&&
R_{(B)}^\al(x; \phi,{\bar g})
\,=\,0,
\nn
\\
&&
R_{(C)}^\al(x; \phi,{\bar g})
\,=\, - \,C^{\sigma}(x)\pa_{\sigma} C^{\alpha}(x),
\nn
\\
&&
R_{({\bar C})}^\al(x;\phi,{\bar g})
\,=\, B^{\alpha}(x).
\eeq
It is easy to see that, in all cases,
$\vp(R^i(x; \phi, {\bar g}))=\vp_i+1$.

There is an important nilpotency property of the BRST
transformations (\ref{A15}), playing a significant role in the proof
of gauge independence of the $S$-matrix on quantum level. To
discuss this property, it is useful to introduce the BRST-operator,
$\widehat{s}$, defined by its action of the fields
\beq
\delta_B \phi^i = \hat{s} \phi^i \mu ,
\qquad
\varepsilon(\hat{s})=1 .
\eeq
Let us formulate the proof of nilpotency using the condensed
DeWitt's notations. Acting twice on all the quantum fields and
using the explicit form (\ref{A15}), we get
\beq
\nonumber
&&
\hat{s}^2 B^{\alpha}\,=\,\hat{s}(\hat{s} B^{\alpha}) =\hat{s}0=0,
\\
\nonumber
&&
\hat{s}^2{\bar C}^{\alpha}\,=\,\hat{s}(\hat{s}{\bar C}^{\alpha})
= \hat{s} B^{\alpha}=0,
\\
\nonumber
&&
\hat{s}^2 C^{\alpha}
= \hat{s}(\hat{s} C^{\alpha})
= -\hat{s}(C^{\sigma}\pa_{\sigma} C^{\alpha})
= C^{\sigma}\pa_{\sigma}(\hat{s} C^{\alpha})
- (\hat{s}C^{\sigma})\pa_{\sigma} C^{\alpha}
\\
\nonumber
&&
\qquad\;\;\,\, =\,- \,
C^{\sigma}\pa_{\sigma}(C^{\rho}\pa_{\rho} C^{\alpha})
+ C^{\rho}\pa_{\rho}C^{\sigma}\pa_{\sigma} C^{\alpha}
\\
\nonumber
&&
\qquad\;\;\,\,
=\,- \,
C^{\sigma}\pa_{\sigma}C^{\rho}\pa_{\rho} C^{\alpha}
- C^{\sigma}C^{\rho}\pa_{\sigma}\pa_{\rho}C^{\alpha}
+ C^{\rho}\pa_{\rho}C^{\sigma}\pa_{\sigma} C^{\alpha}
\\
\nonumber
&&
\qquad\;\;\,\, =\,- \,
C^{\sigma}C^{\rho}\pa_{\sigma}\pa_{\rho}C^{\alpha}=0,
\\
\nonumber
&&
\hat{s}^2h_{\mu\nu}
\,=\, \hat{s}(\hat{s} h_{\mu\nu})
= \hat{s}(R_{\mu\nu\sigma}C^{\sigma})
= (\hat{s}R_{\mu\nu\sigma})C^{\sigma}
+ R_{\mu\nu\sigma}(\hat{s}C^{\sigma})
\\
\nonumber
&&
\qquad\;\;\; \,\,
= \, - \pa_{\sigma}( \hat{s}g_{\mu\nu}) C^{\sigma}
- \pa_{\sigma}g_{\mu\nu}( \hat{s} C^{\sigma})
- ( \hat{s}g_{\mu\sigma}) \pa_{\nu}C^{\sigma}
\\
\nonumber
&&
\qquad\quad
- \,g_{\mu\sigma}\pa_{\nu}(\hat{s}C^{\sigma})
- ( \hat{s}g_{\sigma\nu})\pa_{\mu} C_{\sigma}
- g_{\sigma\nu}\pa_{\mu}(\hat{s}C^{\sigma})
\\
\nonumber
&&
\qquad\quad
= \, \pa_{\sigma}\pa_{\rho}g_{\mu\nu}C^{\rho}C^{\sigma}
+ \pa_{\rho}g_{\mu\nu}\pa_{\sigma}C^{\rho}C^{\sigma}
+ \pa_{\rho}g_{\mu\nu}C^{\sigma}\pa_{\sigma}C^{\rho}
\\
\nonumber
&&\qquad\quad
+\, \pa_{\sigma}g_{\mu\rho}\pa_{\nu}C^{\rho}C^{\sigma}
+ \pa_{\sigma}g_{\mu\rho}C^{\sigma}\pa_{\nu}C^{\rho}
+ \pa_{\sigma}g_{\rho\nu}\pa_{\mu}C^{\rho}C^{\sigma}
+ \pa_{\sigma}g_{\rho\nu}C^{\sigma}\pa_{\nu}C^{\rho}
\\
\nonumber
&&\qquad\quad
+ \,g_{\mu\rho}\pa_{\sigma}\pa_{\nu}C^{\rho}C^{\sigma}
+ g_{\mu\rho}C^{\sigma}\pa_{\sigma}\pa_{\nu}C^{\rho}
+ g_{\rho\nu}\pa_{\sigma}\pa_{\mu}C^{\rho}C^{\sigma}
+ g_{\rho\nu}C^{\sigma}\pa_{\sigma}\pa_{\mu}C^{\rho}
\\
\nonumber
&&\qquad\quad
+ \,g_{\mu\rho}\pa_{\sigma}C^{\rho}\pa_{\nu}C^{\sigma}
+ g_{\mu\rho}\pa_{\nu}C^{\sigma}\pa_{\sigma}C^{\rho}
+ g_{\rho\nu}\pa_{\sigma}\pa_{\mu}C^{\rho}C^{\sigma}
+ g_{\rho\nu}\pa_{\mu}C^{\sigma}\pa_{\sigma}C^{\rho}
\\
\label{nil}
&&\qquad\quad
+ \,g_{\rho\sigma}\pa{\mu}C^{\rho}\pa_{\nu}C^{\sigma}
+ g_{\rho\sigma}\pa_{\nu}C^{\sigma}\pa_{\mu}C^{\rho} =0 ,
\eeq
where the anti-commutativity of ghost fields and the subsequent
relations
\beq
\pa_{\rho}\pa_{\sigma}\phi C^{\sigma}C^{\rho}=0,
\eeq
 valid for an arbitrary function $\phi$, were used.

The BRST invariance of the action $S_{FP}$ can be expressed as
\beq
\label{A18}
\int dx \,\,\frac{\delta_{r} S_{FP}}{\delta\phi^i(x)}\,\,
R^i(x;\phi, {\bar g})\,=\,0.
\eeq
A compact and useful form of the invariance property (\ref{A18})
is called Zinn-Justin equation \cite{Z-J}. To get it, one has to
introduce the set of additional field variables $\phi^*_i(x)$ and
extend the action. The new fields  $\phi^*_i(x)$ are defined to
have Grassmann parities opposite to the corresponding fields
$\,\phi^i(x)$, namely $\,\vp(\phi^*_i) = \vp_i+1$. The extended
action $S=S(\phi,\phi^*,{\bar g})$ has the form
\beq
\label{A19}
S\,\,=\,\,S_{FP}\,+\,\int dx \; \phi^*_i(x) \,R^i(x;\phi, {\bar g}).
\label{extend}
\eeq
It is easy to note that the new fields $\phi^*_i(x)$ play the role
of sources to the BRST generators (\ref{A17}). With this addition,
the relation (\ref{A18}) gets the form of the Zinn-Justin equation
for the action (\ref{A19}),
\beq
\label{A20}
\int dx\,\,\frac{\delta_{r} S}{\delta \phi^i(x)}\,
\frac{\delta_{l} S}{\delta \phi^*_i(x)}\,=\,0.
\eeq
It is worth remembering that using the left and right derivatives in the
last equation is relevant due to the nontrivial Grassmann parities
of the involved quantities.

Let us introduce the terminology of the Batalin-Vilkovisky
formalism \cite{BV,BV1}.
The sources $\,\phi^*_i(x)\,$  are called antifields. Another
fundamental notion is the antibracket for two arbitrary functionals
of fields and antifields, $F=F(\phi,\phi^*)$ and $G=G(\phi,\phi^*)$,
which is defined as
\beq
\label{A22a}
(F,G)\,=\,\int dx \,\bigg(\frac{\delta_{r} F}{\delta \phi^i(x)}
\frac{\delta_{l} G}{\delta \phi^*_i(x)}
-\frac{\delta_{r} F}{\delta \phi^*_i(x)}
\frac{\delta_{l} G}{\delta \phi^i(x)}\bigg),
\eeq
The antibracket obeys the following properties:

1) \ Grassmann parity relations
\beq
\label{A23}
\vp((F, G))=\vp(F)+\vp(G)+1=\vp((G, F));
\eeq

2)  \ Generalized antisymmetry property
\beq
\label{A24}
(F, G)=-(G, F)(-1)^{(\vp(F)+1)(\vp(G)+1)};
\eeq

3)  \ Leibniz rule
\beq
\label{A25}
(F, GH)=(F, G)H+(F, H)G(-1)^{\vp(G)\vp(H)};
\eeq

4) \ Generalized Jacobi identity
\beq
\label{A26}
((F, G), H)(-1)^{(\vp(F)+1)(\vp(H)+1)}
+{\sf cycle} (F, G, H)\equiv 0.
\eeq

In terms of the antibracket, Zinn-Justin equation (\ref{A20}) can
be written as
\beq
\label{A21}
(S,S)\,=\,0,
\eeq
which is the classical master equation of Batalin-Vilkovisky formalism
\cite{BV,BV1}. In what follows, we shall see that this equation
can be generalized to the quantum domain. This generalization is
extensively used to analyse renormalization in QG.

The formulation of the quantum gauge theory starts from the
generating functional of Green functions in the form of functional
integral
\beq
\label{A22}
Z(J,{\bar g})
\,=\,
\int
d\phi\;\exp\Big\{\frac{i}{\hbar}\big[S_{FP}(\phi, {\bar g})+
J\phi\big]\Big\}
\,=\,
\exp\Big\{\frac{i}{\hbar}W(J,{\bar g})\Big\},
\eeq
where $W(J,{\bar g})$ is the generating functional of connected Green
functions. In (\ref{A22}), the DeWitt notations are
\beq
\label{A23a}
J\phi=\int dx\,
J_i(x)\phi^i(x),
\quad
\mbox{where}
\,\,\quad
J_i(x) = \big\{J^{\mu\nu}(x), J^{(B)}_{\alpha}(x),
{\bar J}_\al(x), J_\al(x) \big\}
\qquad
\eeq
are external sources for the fields (\ref{GP}).  The Grassmann
parities and ghost numbers of these sources satisfy the
relations
\beq
\vp(J_i)=\vp(\phi^i),
\qquad
{\rm gh}(J_i)={\rm gh}(\phi^i).
\eeq

Let us start a detailed analysis of the generating functionals and
their gauge dependencies from the generating functional (\ref{A22}).
Later on, to complete the program to prove gauge
invariance of renormalization in QG, we shall introduce a more
general object $Z(J,\phi^*, {\bar g})$, additionally depending on
the antifields $\phi^*$. This extended definition will be given
below, and now we work out a more simple case, just to introduce
the necessary notions in a more transparent way.


As a first step, consider the vacuum functional $Z_{\Psi}({\bar g})$,
that corresponds to the choice of the gauge fixing functional
(\ref{A18}) in the presence  of external metric ${\bar g}$,
\beq
\label{A24original}
Z_{\Psi}({\bar g})
&=&
\int d\phi\;\exp\Big\{\frac{i}{\hbar}\big[S_0({\bar g}+h)
+ \Psi(\phi,{\bar g}) {\hat R}(\phi,{\bar g})\big]\Big\}
\,=\,\exp\Big\{\frac{i}{\hbar}W_{\Psi}({\bar g})\Big\},
\qquad
\eeq
where we introduced the operator
\beq
\label{A24c}
{\hat R}(\phi,{\bar g})=\int dx \,\frac{\delta_r}{\delta\phi^ i(x)}
R^ i(x;\phi,{\bar g})
\eeq
and the fermionic gauge fixing functional $\,\Psi(\phi,{\bar g})$,
\beq
\label{A24b}
\Psi(\phi,{\bar g})=\int dx
\sqrt{-{\bar g(x)}} \;{\bar C}^{\alpha}
\chi_{\alpha}(x;{\bar g},h).
\eeq
Taking into account (\ref{A24c}) and (\ref{A24b}), the expression
in the exponential in the integrand of (\ref{A24original}) is nothing
but $\frac{i}{\hbar}S_{FP}(\phi, {\bar g})$. Thus, the expression
(\ref{A24original}) becomes
\beq
\label{A24a}
Z_{\Psi}({\bar g})
&=&
\int
d\phi\;\exp\Big\{\frac{i}{\hbar}S_{FP}(\phi, {\bar g})\Big\},
\eeq
which is just (\ref{A22}) without the source term in the exponential.

The main advantage of using the fermionic gauge fixing
functional $\,\Psi(\phi,{\bar g})$ is that it is a scalar function
that contains the information about the gauge fixing function
$\chi_{\al}(x;{\bar g},h)$. In particular, the possible change of
the gauge fixing can be explored by evaluating $Z_{\Psi+\de\Psi}$,
that is the modified vacuum functional corresponding to
$\Psi(\phi,{\bar g})+\delta\Psi(\phi,{\bar g})$. Here,
$\de\Psi(\phi,{\bar g})$ is an infinitesimal functional with odd
Grassmann parity. Besides this requirement, $\de\Psi(\phi,{\bar g})$
maybe an arbitrary variation corresponding to the change of
$\chi_{\al}(x;{\bar g},h)$ in Eq.~(\ref{A24b}).

Taking into account (\ref{A24a}), with the new term we get
the vacuum functional with the variation of the gauge fixing,
\beq
\label{A25a}
Z_{\Psi+\delta\Psi}({\bar g})=\int
d\phi\;\exp\Big\{\frac{i}{\hbar}\big[S_{FP}(\phi, {\bar g})+
\delta\Psi(\phi,{\bar g}){\hat R}(\phi,{\bar g})\big]\Big\}.
\eeq

The next step is to make the change of quantum variables
$\phi^{\;\!i}$ in the form of BRST transformations (\ref{A15}),
but with replacing the constant parameter $\mu$ by a functional
$\mu=\mu(\phi,{\bar g})$,
\beq
\label{A26a}
\phi^i (x)\,\,\,\, \longrightarrow\,\,\,\,
\phi^{\prime i}(x)=\phi^i(x) + R^i(x;\phi,{\bar g})
\mu(\phi,{\bar g})
= \phi^i(x) +\Delta \phi^i(x).
\quad
\eeq
In what follows we shall use compact notations for the
generators $R^i(x;\phi,{\bar g}) = R^i(x)$ and for the field
$\mu(\phi,{\bar g})=\mu$. Owing to the linearity of BRST
transformations and the absence of derivatives acting on the
parameter $\mu$, the total action $S_{FP}(\phi, {\bar g})$
remains invariant under
(\ref{A26a}) even for the non-constant $\mu$. It is easy to check
that the Jacobian of transformations (\ref{A26a}) reads \cite{GT}
\footnote{Let us note that the Jacobian of the transformations
(\ref{A26a}) can be evaluated exactly \cite{LL,BLT-FBRST}
but we will not reproduce this calculation here.}
\beq
\label{A26b}
J \,=\,J(\phi, {\bar g})
\,=\,\exp\Big\{\int dx (-1)^{\vp_i}M^{i}_{\;i}(x,x)\Big\},
\eeq
where matrix $M^{i}_{\;j}(x,y)$ has the form
\beq
\label{A26c}
M^{i}_{\;j}(x,y)
\,=\,
\frac{\delta_r \Delta \phi^i(x)}{\delta\phi^j(y)}
\,=\,
(-1)^{\vp_j+1}\frac{\delta_r \mu}{\delta \phi^{j}(y)} R^i(x)
\,-\,(-1)^{\varepsilon_j(\varepsilon_i+1)}
\frac{\delta_l  R^i(x)}{\delta\phi^j(y)}\,\mu.
\quad
\eeq
In the theories of the Yang-Mills type because of the antisymmetry
of the structure constants, there is the following relation:
\beq
\label{A29}
\int  dx \,(-1)^{\vp_i}\,\frac{\de_l R^i(x)}{\de\phi^i(x)}\,=\,0.
\eeq
This relation can be verified using the definitions (\ref{A17}).
Using (\ref{A29}), from (\ref{A26b}) and (\ref{A26c}) it follows that
\beq
\label{A27}
J=\exp\{-\mu(\phi,{\bar g}){\hat R}(\phi,{\bar g})\},
\eeq
where we used the operator (\ref{A24c}).

Choosing the functional $\mu$ in the form
\beq
\label{A26d}
\mu \,=\, \frac{i}{\hbar}\delta\Psi(\phi,{\bar g}),
\eeq
one can observe that the described change of variables in the
functional integral completely compensates the modification in the
expression (\ref{A25a}) compared to the initial formula (\ref{A24a}).
Thus, we arrive at  the gauge independence of the vacuum functional
\beq
\label{A28}
Z_{\Psi}({\bar g})
\,\,=\,\,
Z_{\Psi+\delta\Psi}({\bar g}).
\eeq
The last identity can be written as the vanishing variation of
   the vacuum functionals $Z$ or for $W$, as introduced in
(\ref{A24original}),
\beq
\label{A28a}
\delta_{\Psi}Z({\bar g})\,=\,0
\qquad
\mbox{or}
\qquad
\delta_{\Psi}W({\bar g})\,=\,0.
\eeq
The invariance (\ref{A28}) shows that we can omit the label
$\Psi$ in the definition of the generating functionals (\ref{A22}).

The importance of the invariance (\ref{A28}) is related to the
equivalence theorem by Kallosh and Tyutin \cite{KT}. This QFT
theorem states that if the vacuum functional is invariant, the
$S$-matrix does not depend on the gauge fixing choice (see also
\cite{Costa-Tonin,Kummer}). In the QG case, the existence of the
$S$-matrix requires that the background metric ${\bar g}_{\mu\nu}$
admits asymptotic states. In particular, this requirement is satisfied
if there is a flat Minkowski metric in some extremes, such that, in
these regions, one can form asymptotic $in$ and $out$ states for
the gravitons or, more generally, for the gravitational perturbations
over the flat spacetime. If this condition is satisfied, the invariance
(\ref{A28}) implies that the $S$-matrix in the QG theory does not
depend on the gauge fixing. One can say that if the theory and the
physical situation to which this theory is applied, admit the
construction of the $S$-matrix, the last will be independent on
the choice of the gauge fixing conditions.

It is remarkable that we can make such a strong statement for an
arbitrary model of QG, even without requiring renormalizability
of the theory or the locality of the classical action.  Let us stress
that the statement formulated above is valid only within the
conventional perturbative approach to QFT, or to the QG, as
particular case. On the other hand, the situation may be completely
different in other approaches. The discussion of these situations
(including the asymptotic safety scenario in QG) and further
references can be found in Ref.~\cite{Lavrov-renQG}.

On another hand, in QG gravity the $S$-matrix can be hardly
seen as the object of central interest. One of the reasons is that
the gravitational fields in the asymptotic states should be free
and this cannot be provided in the framework of general relativity.
On the other hand, the main applications of QG are assumed in
cosmology and black hole physics. And in both cases, there are
no direct relations to the scattering problems and the $S$-matrix
does not look the most appropriate object to explore. Taking
these points into account, it may be more advantageous to work
with the effective action, taking control of its non-universality,
i.e., the dependence on the gauge fixing and, more generally, on
the parametrization of quantum fields. Let us now see how this
can be done.

The effective action $\Ga(\Phi,{\bar g})$
is defined by means of Legendre transformation,
\beq
\label{A29a}
\Gamma(\Phi,{\bar g}) \, = \, W(J,{\bar g})\,-\,J_i\Phi^i,
\eeq
where $\Phi=\{\Phi^i\}$ are mean fields and $J_i$ are the solutions
of the equations
\beq
\frac{\delta W(J,{\bar g})}{\delta J_i} \,=\,\Phi^i
\quad
\mbox{and}
\quad
J_i\Phi^i=\int dx \;\!J_i(x)\Phi^i(x).
\eeq
In terms of effective action, the property (\ref{A28a}) means
the on shell independence on the gauge fixing condition, and
can be cast in the form
\beq
\label{A29b}
\delta_{\Psi}\Gamma(\Phi,{\bar g})
\Bigg|_{\frac{\delta\Gamma(\Phi,{\bar g})}{\delta\Phi}=\;\!0}
=\,\,\, 0,
\eeq
i.e., the effective action evaluated on its extremal does not depend
on gauge.

Until now, we did not assume that the background metric may transform
under the general coordinate transformation. This was a necessary
approach, as we mentioned after the definition of the splitting
(\ref{A9}) of the metric into the background and quantum parts.
However, when the effective action is defined, we can perform the
coordinate transformation of the background metric
${\bar g}_{\mu\nu}$, together with the corresponding transformation
for the quantum metric. Taking a variation of the background metric
${\bar g}_{\mu\nu}$ under general coordinate transformations, we get
\beq
\label{A30}
\delta^{(c)}_{\omega}{\bar g}_{\mu\nu}
\,=\,R_{\mu\nu\sigma}({\bar g})\;\!\omega^{\sigma}.
\eeq
Here, the symbol ${(c)}$ indicates that the transformation concerns
the background metric, i.e., is performed in the sector of classical
fields. It is important that this transformation does not lead
neither to the change of the form of the Faddeev-Popov action
(\ref{A11}) nor to the change of the transformation rules for the
auxiliary and ghost fields.

In the quantum fields sector $h_{\mu\nu}$,  the form of the
transformations is fixed by requiring the invariance of the action,
\beq
\label{A32}
\delta^{(q)}_{\omega}h_{\mu\nu}
\,=\,R_{\mu\nu\sigma}(h)\;\!\omega^{\sigma}
\,=\,-\,\omega^{\sigma}\pa_{\sigma}h_{\mu\nu}
- h_{\mu\sigma}\pa_{\nu}\omega^{\sigma}
-h_{\sigma\nu}\pa_{\mu}\omega^{\sigma},
\eeq
where the symbol $(q)$ indicates the gauge transformations
of quantum fields. For the action we have
\beq
\label{A33}
\delta_{\omega}S_0({\bar g}+h)=0,
\qquad
\delta_{\omega}\,=\,
\delta^{(c)}_{\omega}+\delta^{(q)}_{\omega}.
\eeq

With these definitions, for the variation of $Z({\bar g})$ we have
\beq
\label{A31}
\delta^{(c)}_{\omega}Z({\bar g})
&=&
\frac{i}{\hbar}\int d\phi\Big[\delta^{(c)}_{\omega}S_0({\bar g}+h)
+ \delta^{(c)}_{\omega}S_{gh}(\phi,{\bar g})
+ \delta^{(c)}_{\omega}S_{gf}(\phi,{\bar g})\Big]
\nn
\\
&&
\quad
\times \,\,\exp\Big\{\frac{i}{\hbar}S_{FP}(\phi, {\bar g})\Big\}.
\mbox{\qquad}
\eeq
Let us stress that, for a while, we consider only the transformations
of ${\bar g}$, such that $\de^{(q)}$ does not enter the last expression.
At the next stage, using a change of variables in the functional
integral (\ref{A31}) we have to arrive at the relation
$\de^{(c)}_{\om}Z({\bar g})=0$, that is to prove invariance
of $Z({\bar g})$ under the transformations (\ref{A30}).

The gauge fixing action $S_{gf}(\phi,{\bar g})$ depends only on the
three field variables $h_{\mu\nu}$, \ $B^{\alpha}$, and \
${\bar g}_{\mu\nu}$. For $h_{\mu\nu}$ and ${\bar g}_{\mu\nu}$,
the transformation law has been already defined in (\ref{A30}) and
(\ref{A32}). Thus, we need to define the transformation for the
remaining field $B^{\alpha}$. The new rule $\de^ {(q)}_\om B^\al$
should compensate the variation of $S_{gf}(\phi,{\bar g})$
caused by the transformations of ${\bar g}_{\mu\nu}$ and
$h_{\mu\nu}$. The corresponding condition has the form
\beq
\label{A34}
\delta_{\omega}S_{gf}
&=&
\int dx \sqrt{-{\bar g}}\,\,\Big[\big(\delta^{(q)}_{\omega}B^{\alpha}
+\omega^{\sigma}
\pa_{\sigma}B^{\alpha}\big)\chi_{\alpha}({\bar g},h)
\nn
\\
&&
\quad
+ \,\, B^{\alpha}\omega^{\sigma}\pa_{\sigma}\chi_{\alpha}({\bar g},h)
+ B^{\alpha} \delta_{\omega}\chi_{\alpha}({\bar g},h)\Big].
\mbox{\qquad}
\eeq

The transformation of the gauge fixing functions $\chi_{\alpha}$
cannot be defined independently since they are constructed from
the metric and the last transforms according to Eq.~(\ref{A30}).
Thus, the variation of the gauge fixing functions $\chi_{\al}$ has
the form of the vector field transformation (\ref{A5}),
\beq
\label{A35}
\delta_{\omega}\chi_{\alpha}=
-\omega^{\sigma}\pa_{\sigma}\chi_{\alpha}
- \chi_{\sigma}\pa_{\alpha}\omega^{\sigma}.
\eeq
It its turn, the transformation of the auxiliary field $B$ can be
chosen according to the same vector rule (\ref{A6}). This gives
\beq
\label{A37}
\delta^{(q)}_{\omega}B^{\alpha}
= - \omega^{\sigma}\pa_{\sigma}B^{\alpha}
+B^{\sigma}\pa_{\sigma}\omega^{\alpha}\,.
\eeq
It is easy to check that (\ref{A35})  and (\ref{A37})  provide the
desired invariance in  (\ref{A34}),
\beq
\label{A38}
\delta_{\omega}S_{gf}=0.
\eeq

In the same way, one can confirm the  invariance of the ghost action,
\beq
\label{A39}
\delta_{\omega}S_{gh}=0
\eeq
for the vector transformation laws for the ghost fields
${\bar C}^{\alpha}$ and $C^{\alpha}$,
\beq
&&
\delta^{(q)}_{\omega}{\bar C}^{\alpha}(x)
= - \omega^{\sigma}(x)\pa_{\sigma}{\bar C}^{\alpha}(x)
+ {\bar C}^{\rho}\pa_{\rho}\omega^{\alpha}(x),
\nn
\\
&&
\delta^{(q)}_{\omega}C^{\alpha}(x)
= - \omega^{\sigma}(x)\pa_{\sigma}C^{\alpha}(x)
+ C^{\rho}\pa_{\rho}\omega^{\alpha}(x).
\label{A43}
\eeq

All in all, we can conclude that the total Faddeev-Popov
action $S_{FP}$ is invariant
\beq
\label{A40}
\delta_{\omega}S_{FP}=0
\eeq
under the new version of gauge transformations, which
is based on the background transformations of quantum fields
$\phi$ and of ${\bar g}$, i.e., (\ref{A30}),  (\ref{A32}),
(\ref{A37}) and  (\ref{A43}).

As a consequence of (\ref{A40}), important property (\ref{A24a})
and the invariance of the integration measure, the vacuum functional
possesses gauge invariance,
\beq
\label{A44}
\delta_{\omega}Z({\bar g})
\,\,=\,\,
\delta^{(c)}_{\omega}Z({\bar g})
\,\,=\,\,0.
\eeq

As we shall see in what follows, one can use Eq.~(\ref{A44}) to prove
the gauge invariance of an important object called the background
effective action, i.e., the effective action with the switched off mean
fields, \ $\Ga({\bar g})=\Ga(\Phi=0,{\bar g})$. The required feature
can be formulated as
\beq
\label{A45}
\delta^{(c)}_{\omega}\Gamma ({\bar g})=0.
\eeq
Let us note that switching off  the mean quantum fields
$\,\Phi = \{h,\,C,\,{\bar C},\,B\}\,$ requires special care and we
shall see how this should be done. Then, the resulting relation
(\ref{A45}) is one of the main targets of our consideration.

It is useful to start by exploring the off-shell gauge invariance
of the generating functionals of our interest. To this end, it is
useful to present the background transformations  (\ref{A30}),
(\ref{A32}),  (\ref{A37}) and  (\ref{A43}) in the form
\beq
\label{A46}
\delta^{(c)}_{\omega}{\bar g}_{\mu\nu}
\,=\,R_{\mu\nu\sigma}({\bar g})\omega^{\sigma},
\qquad
\delta^{(q)}_{\omega}\phi^i
\,=\,{\cal R}^i_{\sigma}(\phi)\omega^{\sigma},
\eeq
where the generators ${\cal R}^i_{\sigma}(\phi)$ are linear in the
quantum fields $\phi$ and do not depend on the background metric
${\bar g}$.  Utilizing these notations, the general form of the
transformation of an arbitrary functional $\Ga=\Ga(\phi,{\bar g})$
can be written in the form
\beq
\label{A46extra}
\delta_\om \Ga
&=&\,
\delta^{(c)}_{\omega}\Ga
\,+\, \frac{\de_r \Ga}{\de \phi^i}\,{\cal R}^i_\si (\phi) \om^\si.
\eeq

Consider the variation of the generating functional
$Z(J, {\bar g})$ (\ref{A22}) ,
under the gauge transformations of the background metric
\beq
\label{A48}
\delta^{(c)}_{\omega} Z(J, {\bar g})
\,\,=\,\,
\frac{i}{\hbar}\int d \phi \;
\delta^{(c)}_{\omega}S_{FP}(\phi,{\bar g})
\, \exp\Big\{\frac{i}{\hbar}\big[S_{FP}(\phi,{\bar g})
+ J \phi\big]\Big\}.
\eeq
Using the background transformations  in the sector of quantum
fields $\phi$ and taking into account that for the linear change of
variables the Jacobian of this transformation is independent on
the fields, we arrive at the relation
\beq
 \label{A49}
\frac{i}{\hbar}\int d \phi \;
\Big\{
\delta^{(q)}_{\omega}S_{FP}(\phi,{\bar g})
+ J \delta^{(q)}_{\omega}\phi\Big\}\,
 \exp\Big\{\frac{i}{\hbar}\big[S_{FP}(\phi,{\bar g})
 + J\phi\big]\Big\}\,=\,0.
\eeq
On the other hand, from (\ref{A40}) and (\ref{A49}) it follows
that
\beq
\delta^{(c)}_{\omega} Z(J, {\bar g}) &=&
\frac{i}{\hbar}\int d \phi \,\, J_j\,{\cal R}^j_{\sigma}(\phi)
\,\omega^{\sigma} \, \exp\Big\{\frac{i}{\hbar}\big[S_{FP}(\phi,{\bar
g}) + J\phi\big]\Big\}, \nn
\\
&&
=\,\,
\frac{i}{\hbar}J_j{\cal R}^j_{\sigma}
\Big(\frac{\hbar}{i}\frac{\delta}{\delta J}\Big) Z(J, {\bar g})\;\omega^{\sigma}.
\label{A50a}
\eeq

In terms of the generating functional
\ $W=W(J,{\bar g})=-i\hbar\ln Z(J,{\bar g})$ \
of connected Green functions,  the relation (\ref{A50a}) reads
\beq
\label{A51}
\delta^{(c)}_{\omega} W(J, {\bar g})
\,\,=\,\,
J_j\,{\cal R}^j_{\sigma}\,
\Big(\frac{\delta W}{\delta J}\Big)\;\omega^{\sigma},
\eeq
where we used linearity of generators ${\cal R}^i_{\sigma}(\phi)$
with respect to $\phi$. Now we can consider the generating functional
of vertex functions (the effective action),
\beq
 \label{A52}
&&
\qquad\qquad\qquad\qquad
\Ga \,=\,
\Gamma(\Phi, {\bar g})
\,=\, W(J, {\bar g}) -J\Phi,
\\
&&
\mbox{ where}
\qquad
\Phi^j =\frac{\delta_l W}{\delta J_j},
\qquad
\frac{\delta_r \Gamma}{\delta \Phi^j}=-J_j
\qquad
\mbox{and}
\qquad
\delta W=\delta \Gamma .
\qquad
\eeq
Taking the variation of external metric and the mean fields
(\ref{A46}), in terms of $\Ga$, the relation (\ref{A51}) becomes
\beq
 \label{A53}
\delta^{(c)}_{\omega} \Gamma(\Phi, {\bar g})
\,\,=\,\,-\, \frac{\delta_r \Gamma}{\delta \Phi^j}
\,\,{\cal R}^j_{\sigma}(\Phi)
\;\omega^{\sigma}.
\eeq
Finally, taking into account the variations of all fields (\ref{A46})
and using the identity (\ref{A46extra}), we arrive at
\beq
\label{A54}
\delta_{\omega} \Gamma(\Phi, {\bar g})\,\,=\,\,0.
\eeq

It turns out that the relations  (\ref{A53}) and  (\ref{A54}) prove
the fundamental property (\ref{A45}). In order
to see this, one has to note that the generators of quantum fields
(\ref{A32}), (\ref{A43}) and (\ref{A37}) have linear dependence
of these fields. As a result, one meets the following limit for the
generators ${\cal R}^i_{\sigma}(\Phi)$ when the mean fields
are switched off:
\beq
\label{A54b}
\lim_{\Phi \rightarrow 0}{\cal R}^j_{\sigma}(\Phi)=0.
\eeq
This relation shows that $\Ga$ is invariant  under non-deformed
background transformations, i.e, possesses the same invariance as
the classical Faddeev-Popov action.

In the renormalization program based on Batalin-Vilkovisky
formalism, the extended
action is $S=S(\phi,\phi^*,{\bar g})$ defined in (\ref{A19}). The
precursors for the full effective action are the extended generating
functional of Green functions \ $Z=Z(J,\phi^*,{\bar g})$ and of the
connected Green functions \ $W=W(J,\phi^*,{\bar g})$,
\beq
\label{A54a}
Z(J,\phi^*,{\bar g})\,=\,
\int d\phi\;\exp\Big\{\frac{i}{\hbar}\big[S(\phi, \phi^ *,{\bar g})
+ J\phi\big]\Big\}
\,=\,
\exp\Big\{\frac{i}{\hbar}W(J,\phi^*,{\bar g})\Big\}.
\quad
\eeq

Our first purpose is to prove that the effective action satisfies
the quantum version of Eq.~(\ref{A21}). Due to the invariance
of $S_{FP}$ under background fields transformations, the
variation of $S$ takes the form
\beq
\label{A55}
\delta_{\omega}S(\phi,\phi^*,{\bar g})
\,=\,\phi^*_i \delta_{\omega}R^i(\phi,{\bar g}),
\eeq
that shows that the action is gauge invariant on the hypersurface
$\,\phi^*_i=0$.

Using the condensed DeWitt's notation one can write the variations
of the generators $\delta_{\omega}R^i(\phi,{\bar g})$ in the
following compact form:
\beq
\nonumber
&&
\delta_{\om}R^{(h)}_{\,\mu\nu}(\phi,{\bar g})
\,=\,
- \, \om^{\si}\pa_{\si}R_{\mu\nu\la}({\bar g}+h)C^{\la}
- \pa_{\mu}\om^{\si}R_{\si\nu\la}({\bar g}+h)C^{\la}
\nn
\\
&&
\qquad \qquad \qquad \quad
-\,\pa_{\nu}\om^{\si}R_{\mu\si\la}({\bar g}+h)C^{\la},
\nn
\\
&&
\delta_{\om}R_{(B)}^\al (\phi,{\bar g})
\,=\, 0,
\nn
\\
&&
\delta_{\om}R_{(C)}^\al (\phi,{\bar g})
\,=\,
\om^{\si}\pa_{\si}(C^{\la}\pa_{\la}C^{\al})-
C^{\lambda}\pa_{\la}C^{\si}\pa_{\si}\om^{\al},
\nn
\\
&&
\delta_{\om}R_{({\bar C})}^\al(\phi,{\bar g})
\,=\,
-\om^{\si}\pa_{\si}B^{\alpha}+
B^{\si}\pa_{\si}\om^{\alpha}.
\label{A56}
\eeq
The variations $ \delta_{\omega}R^i(\phi,{\bar g})$ are at most
quadratic in the sector of fields $h_{\mu\nu}$ and $C^{\alpha}$
and linear in the field ${\bar C}^{\alpha}$.

Let us now consider the variation of the extended generating functional
$Z(J,\phi^*,{\bar g})$ (\ref{A54a}) under the gauge transformations of
external metric ${\bar g}$,
\beq
\label{A57}
\delta^{(c)}_{\omega}Z(J,\phi^*,{\bar g})
&=&
\frac{i}{\hbar}\int d\phi\,\,
\Big\{
\de^{(c)}_{\om}S_{FP}(\phi,{\bar g})
+ \phi^*_i\de^{(c)}_{\om}R^i(\phi,{\bar g})
\Big\}
\nn
\\
&&
\times\,\,
\exp\Big\{\frac{i}{\hbar}\big[S(\phi,\phi^* {\bar g})+J\phi\big]\Big\}.
\mbox{\qquad}
\eeq
Making the change of variables $\phi^i$ according to (\ref{A32}),
(\ref{A37}), and  (\ref{A43}) in the functional integral and taking
into account the triviality of the corresponding Jacobian, we
arrive at the relation
\beq
 \label{A58}
&&
\frac{i}{\hbar}\int d\phi
 \,\Big\{
 \delta^{(q)}_{\omega}S_{FP}(\phi,{\bar g})
 + \phi^*_i\delta^{(q)}_{\omega}R^i(\phi,{\bar g})
 + J_i\delta^{(q)}_{\omega}\phi^i\Big\}
\nn
\\
&&
\qquad
\qquad
\times\,\,
 \exp\Big\{\frac{i}{\hbar}\big[S(\phi,\phi^* {\bar g})
 + J\phi\big]\Big\}=0.
\mbox{\qquad}
\eeq

Combining Eqs.~(\ref{A57}) and (\ref{A58}) and using the
invariance of $S_{FP}$ (\ref{A40}), we obtain
\beq
 \label{A59}
 \delta^{(c)}_{\omega}Z(J,\phi^*,{\bar g})
&\,\,=\,\,&
\frac{i}{\hbar}\int d\phi \; \Big\{
\phi^*_i \delta_{\omega}R^i(\phi,{\bar g})
+ J_i{\cal R}^i_{\sigma}(\phi)\omega^{\sigma}\Big\}
\nn
\\
&&
\times\,\,
\exp\Big\{\frac{i}{\hbar}\big[S(\phi,\phi^* {\bar g})+J\phi\big]\Big\},
\mbox{\qquad}
\eeq
or, equivalently,
\beq
\label{A60}
\delta^{(c)}_{\omega}Z(J,\phi^*,{\bar g})
&\,\,=\,\,&
\frac{i}{\hbar}\phi^*_i\,
\delta_{\omega}R^i\Big(\frac{\hbar}{i}
\frac{\delta}{\delta J},{\bar g}\Big)\,Z(J,\phi^*,{\bar g})
\nn
\\
&&
+\,\, \frac{i}{\hbar} J_i\,{\cal R}^i_{\sigma}
\Big(\frac{\hbar}{i}\frac{\delta}{\delta J}\Big)
\,Z(J,\phi^*,{\bar g}) \omega^{\sigma}.
\mbox{\qquad}
\eeq

In terms of the generating functional of connected Green functions,
(\ref{A60}) becomes
\beq
\label{A61}
 \delta^{(c)}_{\omega}W(J,\phi^*,{\bar g})
 \,\,=\,\,\,
 \phi^*_i \delta_{\omega}R^i\Big(\frac{\delta W}{\delta J}
 + \frac{\hbar}{i}\frac{\delta}{\delta J},{\bar g}\Big)
\,{\vec 1}
\,+\,
J_i\,{\cal R}^i_{\sigma}\,\Big(\frac{\delta W}{\delta J}\Big)
\, \om^{\si},
\eeq
where the symbol ${\vec 1}$ means that the operator acts on the
unit, ${\vec 1}=1$. In the case of functional derivative, one has
$\frac{\de}{\de \phi}{\vec 1}=0$, but since in many cases the
expressions are non-linear, this is a useful notation.

The extended generating functional of vertex functions (extended
effective action) 
is defined in a standard way. Starting from $W=W(J,\phi^*,{\bar g})$,
introduced in Eq.~(\ref{A54a}), we perform Legendre transformation,
\beq
\label{A63}
\Gamma(\Phi,\phi^ *, {\bar g})
\,\,=\,\,
W(J,\phi^ *,{\bar g})-J\Phi,
\qquad
\Phi^j=\frac{\delta_l W}{\delta J_j},
\qquad
\frac{\delta_r \Gamma}{\delta \Phi^ j}=-J_j.
\eeq
One can easily see that this definition is a generalization of
Eq.~(\ref{A29a}).

As usual for the Legendre transformation, from (\ref{A63})
follows that
\beq
\label{A63double}
\frac{\de_r}{\delta J_k}
\bigg( \frac{\de_l W}{\delta J_i}\bigg)
\,\times\,
\frac{\de_l }{\delta \Phi^i}
\bigg(\frac{\de_r \Ga}{\delta \Phi^j}\bigg)
\,\,=\,\,
-\,\de^k_j.
\eeq

It proves useful to introduce the following notations:
\beq
\label{A66}
\delta_{\omega}{\bar R}^i(\Phi,\phi^*,{\bar g}) =
\delta_{\omega}R^i({\hat \Phi},{\bar g}) {\vec 1},
\qquad
{\hat \Phi}^ j=\Phi^ j
+  i\hbar \big(\Gamma^ {'' -1}\big)^{jk}
\,\frac{\delta_l}{\delta\Phi^ k},
\eeq
where the symbol $\,(\Gamma^ {'' -1})^ {jk}\,$ denotes the matrix
inverse to
\beq
\label{A63extra}
\big(\Gamma''\big)_{ij} \,\,=\,\,
\frac{\de_l }{\delta \Phi^i}
\bigg(\frac{\de_r \Ga}{\delta \Phi^j}\bigg)
\qquad
\mbox{i.e.,}
\qquad
(\Gamma^ {'' -1})^ {ik}(\Gamma^{''})_{kj}=\delta^ i_j.
\eeq

In terms of extended effective action
(\ref{A63}), Eq.~(\ref{A61}) rewrites as
\beq
\label{A64}
\delta^{(c)}_{\omega}\Gamma(\Phi,\phi^ *, {\bar g})
\,=\,-\,\frac{\delta_r \Gamma}{\delta \Phi^ i}\,
{\cal R}^i_{\sigma}(\Phi) \omega^{\sigma}
\,+\,\phi^ *_i\delta_{\omega}{\bar R}^i(\Phi,\phi^*,{\bar g})
\eeq
or, using the relation (\ref{A46extra}), in the form
\beq
\label{A65}
 \delta_{\omega}\Gamma(\Phi,\phi^ *, {\bar g})
 \,\,=\,\,
 \phi^ *_i \,\delta_{\omega}{\bar R}^i(\Phi,\phi^*,{\bar g}) .
\eeq

At this point, we can draw a general conclusion concerning QG
theories in the background field formalism. At the non-renormalized
level, any covariant QG theory has the following general property:
the extended quantum action $S=S(\phi,\phi^ *,{\bar g})$ satisfies
the classical master (Zinn-Justin) equation of the \BV formalism
\cite{BV,BV1}, as we already anticipated in Eq.~(\ref{A21}).
Furthermore,  the extended effective action
$\Ga=\Ga(\Phi,\phi^ *,{\bar g})$ has the same symmetries and
therefore satisfies the same master equation,
\beq
\label{A69}
(\Gamma,\Gamma) \,=\, 0.
\eeq
According to Eq.~(\ref{A55}), the functional
$S=S(\phi,\phi^ *,{\bar g})$ and, according to (\ref{A65}),
also $\Ga=\Ga(\Phi,\phi^ *,{\bar g})$, are invariant under the
background gauge transformations on the hypersurface
\ $\phi^ *=0$,
\beq
\delta_{\omega}S\big|_{\phi^ *=0}\,=\,0,
\qquad
\delta_{\omega}\Gamma\big|_{\phi^ *=0}\,=\,0
\label{69extra}
\eeq
and, more general, satisfy the relations (\ref{A55}) and (\ref{A65}).

\section{Gauge-invariant renormalizability}
\label{sec3}

Up to now, we were considering the non-renormalized generating
functionals of Green functions. The next step is to prove the gauge
invariant renormalizability, which is the property of renormalized
generating functionals. In the framework of \BV formalism, one
can prove the BRST invariant renormalizability. The last means the
preservation of basic equations (\ref{A21}) for the extended action
$\,S(\phi,\phi^*,{\bar g})$ and an identical equation (\ref{A69})
for the extended effective action $\,\Ga(\Phi,\Phi^*,{\bar g})$
after renormalization. The required identities for the classical
and effective renormalized actions have the form
\beq
\label{B3}
(S_R,S_R)=0
\qquad
\mbox{and}
\qquad
(\Ga_R,\Ga_R)=0.
\eeq
Let us remember that the ``classical'' actions $S$ and $S_R$ are
zero-order approximations of the loop expansions in the parameter
$\hbar$ of the bare and renormalized effective actions $\Ga$ and
$\Ga_R$. In this sense,  Eq.~(\ref{A21}) is the zero-order
approximation of Eq.~(\ref{A69}) and what we have to do now is
to extend these two equations to the renormalized quantities $S_R$
and $\Ga_R$. Our strategy will be to explore this extension order
by order in the loop expansion parameter $\hbar$. In this way,
we can prove that the renormalized actions  $S_R$ and $\Ga_R$
obey the gauge invariance property perturbatively.

\subsection{BRST invariant  renormalization}
\label{subsec3.1}

As it was explained above, our next purpose is to consider the
loop expansion in the master equation (\ref{A69}) and arrive at
the conclusion about the coordinate invariance of the $n$-loop
approximation with switched-off sources.
As a first step,  consider the one-loop approximation for
\ $\Ga=\Gamma(\Phi,\Phi^*,{\bar g)}$. To provide the uniformity
of notations we shall use $\Phi^*=\phi^*$  for the antifields.

The  effective action can be presented in the form
\beq
\label{B4}
\Ga \,\,=\,\, \Gamma^{(1)} + {\cal O}({\hbar}^2)
\,\,= \,\,
S \,+\, \hbar \big[{\Gamma}^{(1)}_{div}
\,+\, {\Gamma}^{(1)}_{fin}\big] \,+\, {\cal O}({\hbar}^2),
\eeq
where $S=S(\Phi,\Phi^*,{\bar g})$ is the tree-level action, and
$\Ga^{(1)}_{div}$ and $\Ga^{(1)}_{fin}$ stand for the divergent
and finite parts of the one-loop approximation for $\Ga$.

In the local models of QG, the locality of the divergent part
of effective action is guaranteed by Weinberg's theorem
\cite{Weinberg-1960} (see also \cite{Collins} for an alternative
proof and further references). Let us note, by passing, that there is an
exception from this general rule, corresponding to the theories
with scalar field(s) and with a spontaneous symmetry breaking,
in the presence of the background gravitational field. In this
situation, there may be nonlocal divergences \cite{Tmn-ABL},
however this can be seen as an exception that confirms the
general rule instead of contradicting it.

Furthermore, even if the starting action is nonlocal, the
UV divergences are expected to be described by local functionals.
The physical reason for this is that the high energy domain always
corresponds to the short-distance limit. And in the case of UV
divergences, the energies are infinitely high, hence the distances
should be infinitely short, which does not leave space for the
non-local divergences.


Since there is a special Section of this handbook devoted to the
nonlocal QG models, we will not extend this discussion and just
mention
Refs.~\cite{Tomboulis-97,Tomboulis-2015,Modesto-nonloc,CountGhost}).
It was argued in these works that the UV divergent part of effective
action, for a wide class of nonlocal models of QG, is local, including
the ones with a non-local classical action. Taking this into account,
we can assume that $\Ga^{(1)}_{div}$ is a local functional. The
divergence determines the form of the counterterms of the one-loop
renormalized action, $\,\De S = - \hbar {\Gamma}^{(1)}_{div}$, such
that the renormalized action is
\beq
\label{B5}
S_{1R} \,\,=\,\, S \,-\, \hbar {\Gamma}^{(1)}_{div},
\eeq
which is also a local functional.

Let us substitute the expansion  (\ref{B4}) in  Eq.~(\ref{A69}) and
preserve terms up to the first order in $\hbar$. Thus,
$\,\Ga^{(1)}_{div}\,$ and $\,\Ga^{(1)}_{fin}\,$ satisfy the equation
\beq
0 \,=\, (\Gamma,\Gamma)
&=& (S,S) \,+\, 2\hbar \big(S,\Gamma^{(1)}_{div}\big)
\,+\,
2\hbar \big(S,\Gamma^{(1)}_{fin}\big) \,+\, O(\hbar^2)
\nn
\\
&=&
2\hbar \big(S,\Gamma^{(1)}_{div}\big)
+ 2\hbar \big(S,\Gamma^{(1)}_{fin}\big)
\,+\,O(\hbar^2),
\eeq
where we used the zero-order equation (\ref{A21}). In the first order
in $\hbar$, we have a vanishing sum of the two terms, one of them is
infinite and hence it has to vanish independently on another one.
Therefore
\beq
\label{B6}
\big(S,\; {\Gamma}^{(1)}_{div}\big) \,=\, 0.
\eeq

Let us consider
\beq
\label{B6a}
\big(S_{1R},\;S_{1R}\big)
\,=\,\big(S,S\big)
\,-\, 2\hbar \big(S,{\Gamma}^{(1)}_{div}\big)
\,+\,\hbar^2\big({\Gamma}^{(1)}_{div} ,{\Gamma}^{(1)}_{div}\big).
\eeq
Taking into account (\ref{A21}) and (\ref{B6}), we find the relation
\beq
\label{B7}
(S_{1R},\;S_{1R}) ={\hbar}^2 E_2,
\eeq
where $E_2$ is the known functional. Thus we have shown
that $S_{1R}$ satisfies the classical master equation (\ref{A21})
up to the terms of order $\hbar^2$,
\beq
\label{B8}
\quad E_2\,=\,
\big(\Ga^{(1)}_{div},\,\Ga^{(1)}_{div}\big).
\eeq

Let us now include one-loop corrections to $S$.  The one-loop
effective action ${\Gamma}_{1R}$ can be constructed by adding a
local counterterm to the ${\cal O}(\hbar)$ part of Eq.~(\ref{B4}). As
usual, the counterterm has the divergent part which cancels the
divergence of  ${\Gamma}^{(1)}_{div}$, and the remaining
contribution is finite and typically depends on the renormalization
parameter $\mu$.  The dependence on $\mu$ always repeats the
form of divergence which, as we have seen above, does not violate
the master equation. Thus, we can simply use (\ref{B5}) and assume
that ${\Ga}_{1R}$ is constructed by following the procedure of
quantization described above, with $S$ replaced by $S_{1R}$.

Being constructed in this way, the functional ${\Gamma}_{1R}$ is
finite in the one-loop approximation and satisfies the equation
\beq
\label{B9}
({\Gamma}_{1R},\;{\Gamma}_{1R})
= {\hbar}^2E_2 + \mbox{{\sl O}}({\hbar}^3).
\eeq


Now it is time to consider the next level of the loop expansion.
Consider the one-loop renormalized effective action, taking
into account the ${\cal O}(\hbar^2)$-terms,
\beq
\label{B10}
\Ga_{1R} \,=\,  S \,+\, \hbar
{\Gamma}^{(1)}_{fin}
\,+\, {\hbar}^2 \big[\Ga^{(2)}_{1,div}
\,+\, \Ga^{(2)}_{1,fin}\big]
\,+\, \mbox{{\sl O}}({\hbar}^3).
\eeq
Here $ \Ga^{(2)}_{1,div}$ and $\Ga^{(2)}_{1,fin}$ are divergent
and finite ${\cal O}(\hbar^2)$ parts of the two-loop effective action
constructed on the basis of $S_{1R}$ instead of $S$. The divergent
part ${\Gamma}^{(2)}_{1,div}$ of the two-loop approximation for
${\Gamma}_{1R}$ determines the two-loop renormalization for
$S_{2R}$, according to
\beq
\label{B11}
S_{2R} \,=\, S_{1R} \,-\, {\hbar}^2 {\Gamma}^{(2)}_{1,div}
\eeq
and satisfies the equation
\beq
\label{12} (S,\;
{\Gamma}^{(2)}_{1,div}) \,=\, E_2.
\nonumber
\eeq
As the next step, let us consider
\beq
\label{B13}
(S_{2R},\;S_{2R})={\hbar}^3E_3 + \mbox{{\sl O}}({\hbar}^4).
\eeq
We have found that $S_{2R}$ satisfies the master equations up to
the terms $\hbar^3 E_3$, where
\beq
\label{B14}
E_3=2({\Gamma}^{(1)}_{div},\;{\Gamma}^{(2)}_{1,div}),
\eeq
The effective action $\Gamma_{2R}$ is generated by replacing
$S_{2R}$ into functional integral instead of $S$. Therefore,
$\Gamma_{2R}$  is automatically finite in the two-loop
approximation,
\beq
\nonumber \label{B16} {\Gamma}_{2R}= S + \hbar
{\Gamma}^{(1)}_{fin} + {\hbar}^2{\Gamma}^{(2)}_{1,fin}
 + {\hbar}^3 \big[{\Gamma}^{(3)}_{2,div} +
{\Gamma}^{(3)}_{2,fin}\big] + \mbox{{\sl O}}({\hbar}^4)
\eeq
and satisfies the equation
\beq
\label{B17}
({\Gamma}_{2R},\;{\Gamma}_{2R}) =
{\hbar}^3E_3 + \mbox{{\sl O}}({\hbar}^4).
\eeq

Furthermore, by applying the induction method we find that the
totally renormalized action $S_R$ is given by the expression
\beq
\label{B18}
S_R \,=\,
S \,-\,
\sum_{n=1}^{\infty}{\hbar}^n {\Gamma}^{(n)}_{n-1,div}.
\eeq
We assume that $\Ga^{(n)}_{n-1,div}$ and $\Ga^{(n)}_{n-1,fin}$
are the divergent and finite parts of the $n$-loop approximation for
the effective action, which is already finite in $(n-1)$-loop
approximation,
since it is constructed on the basis of the action $S_{(n-1)R}$.

The action (\ref{B18}) is a local functional and exactly satisfies
the classical  master equation
\beq
\label{B19} (S_R,\;S_R)=0.
\eeq
We have shown that this relation holds in all orders of the loop
expansion. This means that the BRST symmetry is preserved in
the renormalized action $S_R$. Let us note that this corresponds
exactly to the BRST cohomology on local functionals with ghost
number zero \cite{BBH1,BBH2}.

The renormalized effective action ${\Gamma}_R$ is finite in
each order of the loop expansion in the  powers of $\hbar$,
\beq
\label{B20}
{\Gamma}_R = S + \sum_{n=1}^{\infty}{\hbar}^n
{\Gamma}^{(n)}_{n-1,fin},
\eeq
and satisfies the gravitational analog of the Slavnov-Taylor
identities \cite{tHooft-71,Slavnov,Taylor} in Yang-Mills theory
(see also \cite{Weinberg} for the pedagogical introduction),
Thus, the renormalized action $S_R$ and the effective action
${\Gamma}_R$ satisfy the classical master equation and the
gravitational version of Ward (Slavnov-Taylor) identity,
respectively.

\subsection{
Gauge invariance of renormalized background effective action}
\label{subsec3.2}

The next part of our consideration will be to generalize the
transformation relations (\ref{A55}) and (\ref{A65}) for the
renormalized functionals of classical $S_R$ and effective $\Ga_R$
actions. As before, we consider the transformations perturbatively,
starting from the lowest orders. In the one-loop approximation,
from (\ref{A65}) it follows
\beq
\label{B22}
\delta_{\omega}\Gamma(\Phi,\Phi^*,{\bar g})
&\,=\,&
\Phi^*_i\delta_{\omega}R^i(\Phi,{\bar g})
\,+\,
\hbar \Phi^*_i\,
\de_{\om}{\bar R}^{i(1)}_{div}(\Phi,\Phi^*,{\bar g})
\nn
\\
&&
+ \,\,\,
\hbar
\Phi^*_i\delta_{\omega}{\bar R}^{i(1)}_{fin}(\Phi,\Phi^*,{\bar g})
+ O(\hbar^2),
\eeq
where we used the condensed notations (\ref{A66}). In the last
expression,
\beq
\delta_{\omega}{\bar R}^{i(1)}_{div}(\Phi,\Phi^*,{\bar g})
\qquad
\mbox{and}
\qquad
\delta_{\omega}{\bar R}^{i(1)}_{fin}(\Phi,\Phi^*,{\bar g})
\nn
\eeq
are divergent and finite parts of the one-loop approximation
for the gauge transformations
$\delta_{\omega}{\bar R}^{i}(\Phi,\Phi^*,{\bar g})$, correspondingly.
On the other hand, from (\ref{B4}) we have
\beq
\label{B23}
\delta_{\omega}\Gamma(\Phi,\Phi^*,{\bar g})
&\,=\,&
\delta_{\omega}S(\Phi,\Phi^*,{\bar g})+
\hbar \delta_{\omega}\Gamma^{(1)}_{div}
\,+\,
\hbar \delta_{\omega}\Gamma^{(1)}_{fin}
\,+\,O(\hbar^2).
\eeq
The comparison of the two transformations (\ref{B22}) and
(\ref{B23}) tells us that
\beq
\label{B24a}
\delta_{\omega}\Gamma^{(1)}_{div}
&\,=\,&
\Phi^*_i\delta_{\omega}{\bar R}^{i(1)}_{div}(\Phi,\Phi^*,{\bar g}),
\\
\label{B24b}
\delta_{\omega}\Gamma^{(1)}_{fin}
&\,=\,&
\Phi^*_i\delta_{\omega}{\bar R}^{i(1)}_{fin}(\Phi,\Phi^*,{\bar g}).
\eeq

From Eq.~(\ref{B24a}) and the definition (\ref{B5}) follows that the
one-loop renormalized  action \ $S_{1R}=S_{1R}(\Phi,\Phi^*,{\bar g})$
\ transforms according to
\beq
\label{B25}
\delta_{\omega}S_{1R}\,=\, \Phi^*_i\delta_{\omega}R^{i(1)}_{R},
\eeq
where
\beq
\label{B26}
 R^{i(1)}_{R}=R^{i(1)}_{R}(\Phi,\Phi^*,{\bar g})
 = \delta_{\omega} R^{i}(\Phi,{\bar g})
 -\hbar\delta_{\omega}{\bar R}^{i(1)}_{div}(\Phi,\Phi^*,{\bar g}).
\eeq
The last relations mean that the action $S_{1R}$ is invariant under
the background gauge transformations with one-loop deformed gauge
generators $ R^{i(1)}_{R}$ (\ref{B26}) on the hypersurface $\Phi^*=0$.
Furthermore, due to Eq.~(\ref{B25}), functional $\,\Gamma_{1R}\,$
obeys the transformation rule
\beq
 \label{B27}
\delta_{\omega}\Gamma_{1R}
&=&
\Phi^*_i\delta_{\omega} R^{i}
\,+\,
\hbar \Phi^*_i\delta_{\omega}{\bar R}^{i(1)}_{fin}
\nn
\\
&&
+\,\,\hbar^2 \Big[
\Phi^*_i\delta_{\omega}{\bar R}^{i(2)}_{1,div}
+ \Phi^*_i\delta_{\omega}{\bar R}^{i(2)}_{1,fin}\Big]
\,+\, O(\hbar^3),
\qquad
\eeq
where
$\,\delta_{\omega}{\bar R}^{i(2)}_{1,div}
=\delta_{\omega}{\bar R}^{i(2)}_{1,div}(\Phi,\Phi^*,{\bar g})\,$
and
$\,\delta_{\omega}{\bar R}^{i(2)}_{1,fin}
=\delta_{\omega}{\bar R}^{i(2)}_{1,fin}(\Phi,\Phi^*,{\bar g})\,$
are related to
$\,\Gamma^{(2)}_{1,div}$ and $\Gamma^{(2)}_{1,fin}\,$
in Eq.~(\ref{B10}) as
\beq
\label{B28}
\delta_{\omega} \Gamma^{(2)}_{1,div}
=  \Phi^*_i\delta_{\omega}{\bar R}^{i(2)}_{1,div},
\qquad
\delta_{\omega} \Gamma^{(2)}_{1,fin}
= \Phi^*_i\delta_{\omega}{\bar R}^{i(2)}_{1,fin}.
\eeq
Therefore the functional $\,\Gamma_{1R}\,$ is finite in the one-loop
approximation and is invariant under the background gauge
transformations up to the second order in $\hbar$ on the hypersurface
$\,\Phi^*=0$.

Applying the induction method, one can show that the renormalized
functionals $S_R$  and $\Gamma_R$ satisfy the properties\footnote{We
note that these statements are very close to the results concerning
preservation of global symmetries of initial classical action at
quantum level when the effective action of theory under consideration
is invariant under deformed global transformations of all fields
\cite{BL}.}
\beq
\label{B27a}
\delta_{\omega}S_{R}= \Phi^*_i\,\delta_{\omega}R^i_R \,,
\qquad
\quad
\delta_{\omega}\Gamma_{R}
= \Phi^*_i\,\delta_{\omega}{\bar R}^i_R,
\eeq
where
\beq
\label{B28a}
\delta_{\omega}R^i_R
&\,=\,&
\delta_{\omega} R^{i}
\,-\, \hbar\delta_{\omega}{\bar R}^{i(1)}_{div}
\,-\,\hbar^2\delta_{\omega}{\bar R}^{i(2)}_{1,div}\,-\,\cdots ,
\\
\label{B29}
\delta_{\omega}{\bar R}^i_R
&\,=\,&
\delta_{\omega} R^{i}
\,+\,\hbar\delta_{\omega}{\bar R}^{i(1)}_{fin}
\,+\,\hbar^2\delta_{\omega}{\bar R}^{i(2)}_{1,fin}\,+\,\cdots .
\eeq
It is important that $\,\delta_{\omega}{\bar R}^i_R$,
defined in (\ref{B29}), are finite.

As it was discussed above, the UV divergences in any models of QG
are local, even for the nonlocal QG theories
\cite{Tomboulis-97,Modesto-nonloc,CountGhost}.
As a consequence, the quantities $\,\de_\om R^i_R$,
defined in (\ref{B28a}), are local while in both local and non-local
models of  QG, including the proper transformations $\de_\om$.
Then, an important consequence of the results  (\ref{B27a}) is that
the renormalized functionals
$\,S_R(\Phi,{\bar g})=S_R(\Phi,\Phi^*=0,{\bar g})\,$ and
$\,\Ga_R(\Phi,{\bar g})=\Ga_R(\Phi,\Phi^*=0,{\bar g})\,$
obey the symmetry
\beq
\label{B30}
\de_\om S_R(\Phi,{\bar g}) \,=\, 0,
\qquad
\de_\om\Ga_R(\Phi,{\bar g}) \,=\, 0.
\eeq
These are the same transformations as for the non-renormalized
functionals $S(\Phi,{\bar g})=S_{FP}(\Phi,{\bar g})$ and
$\Ga(\Phi,{\bar g})$ in (\ref{A40}) and (\ref{A54}), respectively.
Therefore, from (\ref{B30}) follows the invariance property of the
renormalized background functionals
$\,S_R({\bar g})=S(\Phi=0,{\bar g})\,$ and
$\,\Ga({\bar g})=\Ga(\Phi=0,{\bar g})\,$ under general coordinate
transformations of external background metric $\,{\bar g}_{\mu\nu}$,
\beq
\label{B31}
\delta^{(c)}_\om S_R({\bar g})=0,
\qquad
\delta^{(c)}_{\omega}\Gamma_R({\bar g})=0.
\eeq
It is easy to see that these invariances repeat exactly the
symmetry properties of initial action $\,S_0({\bar g})\,$  and
$\,\Ga({\bar g})\,$ in (\ref{A44}).


\section{A short historical review and more special notes}
\label{sec4}

In order to understand better the relevance of the results described
above, let us start by presenting a short historical review of the
subject. The first proof of the gauge invariant renormalizability
in QG was given by Stelle in the famous 1977 paper \cite{Stelle77}.
Despite the considerations in this paper were done for the simplest
renormalizable model of quantum gravity with four derivatives,
most of the analysis is sufficiently general and can be applied,
also, to other covariant models of QG. After that, there were
further publications devoted to the invariant renormalizability in QG
\cite{VorTyu82,VorTyu84,Barv,Lavrov-renQG}, where the main
conclusion about the diffeomorphism invariant renormalizability
has been confirmed with different degrees of generality and using
formally different methods, regardless all of the proofs are based
on the BRST symmetry.

Since QG is a particular example of gauge theory, the program of
exploring invariant renormalizability cannot be separated from the
works done in the framework Yang-Mills theories. The most
significant achievement in this respect was the demonstration of
BRST-invariant renormalizability in the theories which may be not
renormalizable by power counting\footnote{The gauge invariant
renormalizability,
independent on the power counting, is especially important for
the effective QFT \cite{Weinberg} and, in particular, for effective
approach to QG. The interested reader can consult a specially
devoted to effective QG Section of the present Handbook for
further details. }. In 1982 it was formulated the first proof for the
general gauge theories \cite{VLT82}, based on the Batalin-Vilkovisky
formalism \cite{BV,BV1}. The approach employed in  \cite{VLT82}
assumed the regularization procedure respecting the gauge invariance
of initial classical action and providing zero volume divergences,
i.e., with the condition $\,\delta(0)=0$ satisfied.
The proof
is valid for any boundary condition related to an initial gauge invariant
action and for arbitrary choice of gauge fixing functions. Furthermore,
the renormalization procedure of \cite{VLT82} can be described in
terms of anticanonical transformations (for recent developments, see
\cite{BLT-EPJC, BL-BV}) which are defined as transformations
preserving the antibracket, using the terminology of the standard
review paper \cite{GPS}.

An alternative, albeit very close, approach to prove the BRST
invariant renormalization of general gauge theories
\cite{Gomis-Weinberg}, is based on the use of cohomologies of
nilpotent BRST operator associated with the adjoint operation of
the antibracket of the action $S$ with an arbitrary functional $F$,
${\hat s}F=(S,F)$ \cite{BBH1,BBH2}. The detailed description
of this approach can be found in \cite{GPS} and in the chapter
17.3 of Vol. II of the book \cite{Weinberg}. The disadvantage
of this approach is that it does not directly extends to the
background field method.

On the other hand, the background field formalism
\cite{DeW,AFS,Abbott} represents a powerful approach to study
quantum properties of gauge theories, allowing to keep the gauge
invariance (general covariance, for QG), at all stages of the loop
expansion, including in practical calculations.  From the viewpoint
of the quantization of gauge systems the background field method
corresponds to the special choice of a boundary condition and to the
special choice of gauge fixing functions. However, since the
background field method requires the presence of an ``external''
field in the course of the Lagrangian quantization, this formalism
should be considered as a very special case which requires special
care. Indeed, this special case attracted a great deal of attention
recently \cite{Barv,BLT-YM,FT,Lav,BLT-YM2}.

A few more special notes on the structure of the gauge algebra
underlying a given classical system are in order. Already the
quantization of gauge systems using the Faddeev-Popov approach
\cite{FP}, should be applied to the theories when the gauge algebra
is associated with a Lie group, as otherwise, the quantization may
give wrong results. The reason is that, in these, more complicated
cases,  the gauge algebra may be reducible/open \cite{Town,dWvH} or
structure functions may depend on the fields of the initial gauge
theory, as it is happens in supergravity \cite{FvNF,DZ,Nie1,Kal}. In
these cases, the symmetry transformations leave the action
invariant, but do not form a gauge group. Thus, the quantization of
gauge theories requires taking into account not only the invariance
of the action, but also such important aspects as open/closed
algebras, the presence of reducible generators, and so on. The
quantization of these complicated theories is possible using
different types of ghosts, antighosts, ghosts for ghosts
(Nielsen-Kallosh ghosts etc.)
\cite{dWvH,Nie1,Kal,Nie2,HKO,DeAGM,FrS,FrT} (for recent applications
to quantization of reducible gauge theories see \cite{O,KNO1,KNO2}).
The most general (and the
unique completely self-consistent) approach to the problem of
covariant quantization, which summarized all these approaches, was
proposed by Batalin and Vilkovisky \cite{BV,BV1}. The
Batalin-Vilkovisky formalism gives the rules of quantization of
the general gauge theories which may be characterized by
open/reducible gauge algebras, with structure functions depending on
the fields of the initial action. The Batalin-Vilkovisky formalism is
not only a powerful quantization method, as it also enables us to
explore various complicated subjects and issues in gauge theories
\cite{VLT82,Gomis-Weinberg,Schwarz,Barnich,BH,BHW,BLT-EPJC,BL-2021}.
And among these important applications, the renormalization of the
quantum gauge theories of different types is one of the main
problems. Coming back to gravity, in the next sections we will see
that QG belongs to the theories of the Yang-Mills type, i.e., it
admits the traditional Faddeev-Popov approach and enables one to
prove the gauge-invariant renormalizability \cite{Stelle77} (see
also \cite{VorTyu84}). However, using the Batalin-Vilkovisky
formalism, we can make
this proof  more elegant and simple, as we shall see in the rest
of this review.

\section{On the gauge fixing in the higher derivative models}
\label{sec5}

Eqs.~(\ref{B27a}) show that with the antifields switched off, i.e.,
with $\Phi^*=0$, both the renormalized classical action $S_R$ and
effective action $\Ga_R$ are gauge invariant quantities. In
particular, this means that if we restrict the consideration by the
standard non-extended generating functional of the Green functions,
that means, without introducing sources for the ghosts $\,C$,
$\,{\bar C}$  and for the auxiliary field $\,B$, the effective action
will be metric-dependent and generally covariant functional. This
statement concerns both divergent and finite parts of renormalized
effective action. In what follows we discuss how the gauge invariant
renormalizability can be applied to the renormalization of the
models of QG. The rest of this section partially repeats some part
of the previous Chapter~\cite{BackQG}, but we shall discuss the
subject from a different perspective.

The use of the power counting arguments in QG models
may be especially simple if the following two conditions
are satisfied:

{\it i)} The propagator of the gravitational
field should be homogeneous in the powers of momenta.
This means, in particular, that the free equations for different
modes of the gravitational field (tensor, vector and scalar)
are of the same order in derivatives after the Faddeev-Popov
procedure.

{\it ii)} The propagator of FP ghosts should have the same
powers  of momenta as all modes of the gravitational field.

As we explained in detail in the previous Chapter, these
conditions are fulfilled if the gauge fixing term and the
ghost action have the form
\beq
S_{gf} &=& \int d^4 x\sqrt{-g}\,\chi^\al\,Y_{\al\be}\,\chi^\be ,
\label{modGF}
\\
S_{gh} &=& \int d^4x\sqrt{-g}\,{\bar C}^\al\,Y_{\al\be}
\,M^\be_{\,\la}\,C^\la ,
\label{modGH}
\eeq
where, according to (\ref{A11a}) and  (\ref{A11b}),
\beq
M^\be_{\,\la} &=&
H_{\be}^{\rho\si}(x,y;{\bar g},h)
R_{\rho\si\la}(y,z;{\bar g}+h).
\label{Mghost}
\eeq
The choice of the weight operator $Y_{\al\be}$ should be done in
such a way that the total amount of derivatives in the expressions
(\ref{modGF}) and (\ref{modGH}) be the same as in the action of
the model of quantum gravity under consideration. For instance, in
the QG based on general relativity $Y_{\al\be}= \th g_{\al\be}$,
where $\th$ is a constant gauge fixing parameter. In case of the
fourth order gravity one has to take  \cite{Stelle77,julton,frts82}
\beq
Y_{\al\be} &=&
\th_1 \de_{\al\be}\cx + \th_2 \na_\al \na_\be + \th_3 R_{\al\be}
+ \th_4 \de_{\al\be}R,
\label{Y4D}
\eeq
where $\th_{1,2,3,4}$ are gauge fixing constants. In the case of
six-derivative superrenormalizable gravity model \cite{highderi}
$\th_{1,2,3,4}$ should be linear functions of d'Alembertian
operator $\,\cx$, plus the
possible linear in curvature tensor  terms, for the eight-derivative
QG the parameters $\th_{1,2,3,4}$ become quadratic functions
of $\cx$, etc.

An important question is how to incorporate the modified gauge fixing
and ghost actions (\ref{modGF}) and (\ref{modGH}) into the proof of
gauge invariant renormalizability which we developed in Sec. \ref{sec3}.
The simplest possibility in this direction is to remember  that the
effective action in the superrenormalizable QG theories with
more than four derivatives does not depend on the gauge fixing
\cite{highderi}. Thus the scheme based on the weight function
(\ref{Y4D}) with $\th_{1,2,3,4}$ linear functions of $\cx$, does
not affect the loop corrections. This argument looks convincing,
but let us present some extra details and refer to
\cite{Lavrov-renQG} for further discussions.

Consider $\chi_{\alpha}=\chi_{\alpha}(x;{\bar g}, h)$ being a
standard gauge fixing function used in previous sections. We can
introduce the set of two  differential operators, $Y_{\al}^{\,\be}$
and $Y_{1\al\be}$. These weight operators must have the structure
of tensor fields of types $\,(1,1)$ and $(0,2)$, respectively, and
can not depend on the quantum metric $h_{\mu\nu}$,
\beq
\label{Y1}
&& Y_{\alpha}^{\beta}(x,y)
\,=\, Y_{\alpha}^{\beta}(x,y;{\bar g}, {\bar \square})
\qquad
\mbox{and}
\qquad
Y_{1\alpha\beta}(x,y)
\,=\,Y_{1\alpha\beta}(x,y;{\bar g}, {\bar \square}).
\quad
\eeq
The next step is to modify the gauge fixing functions $\,\chi_\al$,
by the following rule:
\beq
\chi^{mod}_{\alpha}(x;{\bar g},h, B)
\,=\,
\int dy\, \Big[
Y_{\alpha}^{\be}(x,y;{\bar g}, {\bar \cx})
\,\chi_{\beta}(y;{\bar g}, h)
\,+\,
\frac{1}{2}Y_{1\alpha\beta}(x,y;{\bar g}, {\bar \square})\,
B^{\beta}(y)\Big]
\qquad
\label{2}
\eeq
and construct the corresponding gauge fixing functional,
\beq
\label{3}
\Psi^{mod}(\phi,{\bar g})
\,=\,
\int dx  \sqrt{-{\bar g}}\,\,{\bar C}^{\alpha}(x)\,
\chi^{mod}_{\alpha}(x;{\bar g},h, B).
\eeq
According to what we previously learned, the transformation law of
$\chi^{mod}_{\alpha}$ coincides with the transformation rule of
tensor fields of type $(0,1)$. Then the modified Faddeev-Popov
action is constructed in the standard manner, using the generator
of BRST transformations, \ ${\hat R}(\phi,{\bar g})$,
\beq
\label{4}
S_{FP}^{mod}(\phi,{\bar g})
\,=\,
S_0({\bar g}+h)\,+\,\Psi^{mod}(\phi,{\bar g})
\,{\hat R}(\phi,{\bar g}).
\eeq
The explicit form of the second term in the right-hand side
of (\ref{4}) is
\beq
\nonumber
\Psi^{mod}(\phi,{\bar g}){\hat R}(\phi,{\bar g})
&=& \int dxdydz du\sqrt{-{\bar g}(x)}\;{\bar C}^{\alpha}(x)
Y_{\alpha}^{\beta}(x,u;{\bar g}, {\bar \square})
\label{5}
\\
&&
\times \,\,
H^{\gamma\sigma}_{\beta}(u,y;{\bar g},h)
R_{\gamma\sigma\rho}(y,z;{\bar g}\!+\!h)C^{\rho}(z)
\nn
\\
&&
+ \, \int dxdy  \sqrt{-{\bar g}(x)}
\Big[B^{\alpha}(x)Y_{\alpha}^{\beta}(x,y;{\bar g},
{\bar \square})\chi_{\beta}(y;{\bar g}, h)
\nn
\\
&&
+ \,
\frac{1}{2}B^{\alpha}(x)
Y_{1\alpha\beta}(x,y;{\bar g}, {\bar \square})B^{\beta}(y)\Big].
\nn
\eeq
The first term in the {\it r.h.s.} of the last formula is exactly
(\ref{modGH}) with (\ref{Mghost}). As the  transformation rules
for the terms in the Faddeev-Popov action depend only on the type
of the tensor fields, all the main statements of the previous
sections remain valid for the new choice of the gauge fixing
functions (\ref{2}).

To deal only with the problem of the homogeneity of the
propagator of the quantum metric $h_{\mu\nu}$, consider a
special choice of the weight operator
\beq
&&
Y_{1\al\be}(x,y)
\,=\, {\bar g}_{\al\ga}(x)\,(Y^{-1})^{\ga}_{\be}(x,y),
\nn
\\
\mbox{where}
\quad
&&
\int dz \,\,
Y^{\gamma}_{\alpha}(x,z)\,(Y^{-1})_{\gamma}^{\beta}(z,y)
\,=\, \delta_{\alpha}^{\beta}\delta(x-y),
\nn
\\
&&
\,\,\quad
Y_\al^{\,\be}(x,y)\,=\,Y_{\al}^{\,\be}(x;{\bar g}, {\bar \square})
\,\de(x-y).
\eeq
Integrating over the fields
$B^{\alpha}$ in the functional integral defines the generating functional
of Green functions it terms of integration over fields
${\bar C}^\al$, $C^\al$ and $h_{\mu\nu}$.
As a results we obtain the functional determinant  that is equal to
\beq
\Big[{\rm Det}\;Y_{\alpha}^{\beta}(x,y)\Big]^{1/2},
\label{weight det}
\eeq
independent on the variables of integration. It is worth noting
that the factor (\ref{weight det}) is well-known in higher
derivative QG models \cite{frts82,highderi,MRS}.

After all, to introduce a nontrivial weight operator we need
to replace
\beq
&&
\Psi^{mod}(\phi,{\bar g}){\hat R}(\phi,{\bar g})+
\int dx \sqrt{-{\bar g}(x)} \;J_{\alpha}^{(B)}(x)B^{\alpha}(x)
\eeq
by the more complicated expression
\beq
&&
\int \, dxdydz\sqrt{-{\bar g}(x)}\;{\bar C}^{\alpha}(x)
 \,Y_{\alpha}^{\beta}(x;{\bar g}, {\bar \square}) \,
H^{\ga\si}_{\be}(x,y;{\bar g},h)
 \, R_{\ga\si\rho}(y,z;{\bar g}\!+\!h) \,C^{\rho}(z)
\nn
\\
&&
- \, \frac{1}{2}\!\int dx  \sqrt{-{\bar g}(x)}\;J^{(B)\alpha}(x)
Y_{\alpha}^{\beta}(x;{\bar g}, {\bar \square})J^{(B)}_{\beta}(x)
- \int dx  \sqrt{-{\bar g}(x)}\;J^{(B)}_{\alpha}(x)\;\chi^{\alpha}(x;{\bar g},h)
\nn
\\
&&
-\, \frac{1}{2}\!\int dx  \sqrt{-{\bar g}(x)}\;\chi^{\alpha}(x;{\bar g},h)
Y_{\alpha}^{\beta}(x;{\bar g}, {\bar \square})\chi_{\beta}(x;{\bar g},h),
\label{modPsi}
\eeq
where the notations
\beq
&&
\chi^{\alpha}(x;{\bar g},h)
\,=\,
{\bar g}^{\alpha\beta}(x)\chi_{\beta}(x;{\bar g},h),
\nn
\\
&&
J^{(B)\alpha}(x) \,=\, {\bar g}^{\alpha\beta}(x)J^{(B)}_{\beta}(x)
\eeq
are used. The second term in the expression (\ref{modPsi}) is exactly
what is needed for the homogeneity condition (\ref{modGF}).
The total action appearing after integration over fields $B^{\alpha}$ is
the  sum of the action $S_0({\bar g}+h)$ plus the first and the fourth
terms in (\ref{modPsi}). This total action is invariant under the BRST
transformations, the last are now recast in the form
\beq
\nonumber
&&\delta_B h_{\mu\nu}(x)
=\int dy R_{\mu\nu\alpha}(x,y;{\bar g}+h)C^\al(y)\mu,
\\
\label{BRST-B}
&&
\delta_B C^{\alpha}(x) =-C^{\sigma}(x)\pa_{\sigma}
C^{\alpha}(x)\mu,
\\
&&
\delta_B {\bar C}^{\alpha}(x)
= -\chi^{\alpha}(x;{\bar g},h)\mu .
\nn
\eeq
In this form, the BRST symmetry also enables us to use all the basic
properties to explore the renormalization of quantum theory, as one
can verify by inspecting the considerations in Sec.~\ref{sec2} and
Sec.~\ref{sec3} (see also \cite{Stelle77}). However, there is a
price to pay for the possibility to work with the functional
integral over the smaller number of integration variables. One can
note that the nilpotency property of the BRST transformations is
lost in the new version of BRST transformations (\ref{BRST-B}).
On another hand, in the context of gauge invariant renormalizability,
the nilpotency of the BRST transformation does not play a critical
role. For this reason, we can freely switch between the two versions
of BRST, according to our convenience.

As the problem of homogeneity and introduction of
(\ref{modGF}) and (\ref{modGH}) has been solved, we
are in a position to review the power counting and classify the
models of quantum gravity, as it was explained in detail in the
previous Chapter \cite{BackQG} (see also \cite{Lavrov-renQG} and
\cite{OUP}). Thus, the important understanding of the main aspects
of renormalization and effective approaches in perturbative QG gains
a reliable basis after we get a mathematically solid proof of the
gauge invariant renormalizability of the classically covariant
models of gravity.

\section{Conclusions}
\label{secC}

The general proof of diffeomorphism invariant renormalization
in QG, independent of the renormalizability by power counting,
is relevant for several important reasons. In particular, such a
proof provides a solid basis for the low-energy effective approach
in QG, that means making practical calculations in the low-energy
sector of the theory, even in non-renormalizable models. In this
case, it is necessary to be sure that the UV divergences are
subtracted by local covariant counterterms that do not affect the
physics in the IR \cite{Weinberg}.

The main advantages of the approach of \cite{Lavrov-renQG}
which we reproduced and reviewed here, is related to the compact
description of the variation of extended effective action under the
gauge transformations of all fields used in the background field
formalism. The derived form of these variations can be applied to an
arbitrary QG theory, respecting the diffeomorphism invariance. The
variation shows an invariance of the effective action when the
antifields (sources for the BRST generators) are switched off.

After switching off the mean field of quantum metric, Faddeev-Popov
ghosts, auxiliary field, and antifields, the divergent part of effective
action possess general covariance, and this important property holds
in all orders of the perturbative loop expansion.
This statement holds in all covariant models of QG, including the
ones with higher derivatives and even for the non-local models.
Starting from covariance and using power counting and locality
of the counterterms one can easily classify the models of QG into
non-renormalizable, renormalizable, and superrenormalizable versions.
One of the extensions of the analysis performed above is an
extension to the non-linear gauges \cite{LLSh}, but this part is
beyond the scope of the present review.

\section*{Acknowledgements}
\label{secAck}

The work of P.M.L. is partially supported by the  the Ministry of
Education of the Russian Federation, under the project No.
FEWF-2020-0003.
This work of I.L.Sh. is partially supported by Conselho Nacional de
Desenvolvimento Cient\'{i}fico e Tecnol\'{o}gico - CNPq (Brazil),
the grant 303635/2018-5 and by Funda\c{c}\~{a}o de Amparo \`a
Pesquisa de Minas Gerais - FAPEMIG, the project PPM-00604-18;
and by the Ministry of Education of the Russian Federation, under
the project No. FEWF-2020-0003.


\end{document}